\documentclass[preprint2]{aastex} 
\usepackage{natbib}
\usepackage{graphicx} 
\usepackage{gensymb}
\usepackage{txfonts}
\usepackage{rotating}
\usepackage{textcomp}
\usepackage{url}
\usepackage{hyperref}
\hypersetup{
    colorlinks=false,
    pdfborder={0 0 0},
}
\makeatletter 

\begin
{document}
   \title{The Herschel Comprehensive (U)LIRG Emission Survey (HerCULES): CO Ladders, fine structure lines, and neutral gas cooling \footnote{{\it Herschel} is an ESA space observatory with science instruments provided by European-led Principal Investigator consortia and with important participation from NASA.}}
\author{M.~J.~F.~Rosenberg\altaffilmark{1}, P.~P. van der Werf\altaffilmark{1}, S. Aalto\altaffilmark{2}, L. Armus\altaffilmark{3},V. Charmandaris\altaffilmark{4,23,24}, 
T. D\'iaz-Santos\altaffilmark{3}, A.~S. Evans\altaffilmark{5,22}
J. Fischer\altaffilmark{6}, Y. Gao\altaffilmark{7}, E. Gonz{\'a}lez-Alfonso\altaffilmark{8}, T.~R. Greve\altaffilmark{9}, A.~I. Harris \altaffilmark{10},
C. Henkel\altaffilmark{11,26}, F.~P. Israel\altaffilmark{1},  K.~G. Isaak\altaffilmark{12}, C. Kramer\altaffilmark{13},
R. Meijerink\altaffilmark{1}, D.~A. Naylor\altaffilmark{14}, D.~B. Sanders\altaffilmark{15},
H.~A. Smith\altaffilmark{16}, M. Spaans\altaffilmark{17}, L. Spinoglio\altaffilmark{18}, G.~J. Stacey\altaffilmark{19}, I. Veenendaal\altaffilmark{14}, S. Veilleux \altaffilmark{10,25},
F.~Walter\altaffilmark{20}, A. {Wei{\ss}}\altaffilmark{11}, M.~C. Wiedner\altaffilmark{24}, M.~H.~D. van der Wiel\altaffilmark{14}, E.~M. Xilouris\altaffilmark{4}}

\altaffiltext{1}{Leiden Observatory, Leiden University, P.O. Box 9513, 2300 RA Leiden, The Netherlands \email{rosenberg@strw.leidenuniv.nl}}
\altaffiltext{2}{Department of Earth and Space Sciences, Chalmers University of Technology, Onsala Observatory, 43994 Onsala, Sweden}
\altaffiltext{3}{Spitzer Science Center, California Institute of Technology, MS 220-6, Pasadena, CA 91125, USA}
\altaffiltext{4}{Institute for Astronomy, Astrophysics, Space Applications \& Remote Sensing, National Observatory of Athens, P. Penteli, 15236 Athens, Greece}
\altaffiltext{5}{Department of Astronomy, University of Virginia, P.O. Box 400325, Charlottesville, VA 22904, USA}
\altaffiltext{6}{Naval Research Laboratory, Remote Sensing Division, 4555 Overlook Ave SW, Washington, DC 20375, USA}
\altaffiltext{7}{Purple Mountain Observatory, Chinese Academy of Sciences (CAS), 2 West Beijing Road, Nanjing 210008, China}
\altaffiltext{8}{Universidad de Alcal{\'a}, Departamento de Física y Matem{\'a}ticas, Campus Universitario, 28871 Alcal{\'a} de Henares, Madrid, Spain}
\altaffiltext{9}{Department of Physics and Astronomy, University College London, Gower Street, London WC1E 6BT, UK}
\altaffiltext{10}{Department of Astronomy, University of Maryland, College Park, MD 20742, USA}
\altaffiltext{11}{Max-Planck-Institut f{\"u}r Radioastronomie, Auf dem H{\"u}gel 16, Bonn, D-53121, Germany}
\altaffiltext{12}{Scientific Support Office, ESTEC/SRE-S,Keplerlaan 1, NL-2201 AZ, Noordwijk, The Netherlands}
\altaffiltext{13}{Instituto Radioastronom{\'i}a Milim{\'e}trica (IRAM), Av. Divina Pastora 7, Nucleo Central, 18012 Granada, Spain}
\altaffiltext{14}{Institute for Space Imaging Science, Department of Physics and Astronomy, University of Lethbridge, Lethbridge, AB, Canada}
\altaffiltext{15}{Institute for Astronomy, 2680 Woodlawn Drive, University of Hawaii, Honolulu, HI 96822}
\altaffiltext{16}{Harvard-Smithsonian Center for Astrohpysics, 60 Garden Street, Cambridge, MA 02138, USA}
\altaffiltext{17}{Kapteyn Astronomical Institute, P.O. Box 800, NL-9700 AV Groningen, The Netherlands}
\altaffiltext{18}{Istituto di Astroﬁsica e Planetologia Spaziali, INAF, Via Fosso del Cavaliere 100, I-00133 Roma, Italy}
\altaffiltext{19}{Department of Astronomy, Cornell University, Ithaca, NY 14853, USA}
\altaffiltext{20}{Max-Planck Institut fur Astronomie, K{\"o}nigstuhl 17, D-69117 Heidelberg, Germany}
\altaffiltext{22}{National Radio Astronomy Observatory, 520 Edgemont Road, Charlottesville, VA 22903, USA}
\altaffiltext{23}{University of Crete, Department of Physics, Heraklion 71003, Greece}
\altaffiltext{24}{Observatoire de Paris, LERMA, 61 Avenue de l’Observatoire, 75014 Paris, France}
\altaffiltext{25}{Joint Space-Science Institute, University of Maryland, College Park, MD 20742}
\altaffiltext{26}{Astronomy Department, King Abdulaziz University, P.O. Box 80203, Jeddah 21589, Saudi Arabia}

\begin{abstract}
    (Ultra) Luminous Infrared Galaxies ((U)LIRGs) are objects characterized by their extreme infrared (8-1000 $\mu$m) 
    luminosities ($L_{LIRG}>10^{11} $L$_\odot$ and $L_{ULIRG}>10^{12}$ L$_\odot$). The Herschel Comprehensive ULIRG Emission Survey (HerCULES; PI van der Werf) presents a representative flux-limited sample
    of 29 (U)LIRGs that spans the full luminosity range of these objects (10$^{11}\leq L_\odot \geq10^{13}$).  With the \emph{Herschel Space Observatory}, we
    observe [CII] 157 $\mu$m, [OI] 63 $\mu$m, and [OI] 145 $\mu$m line emission with PACS, CO J=4-3 through J=13-12, [CI] 370 $\mu$m, and [CI] 609 $\mu$m 
    with SPIRE, and low-J CO transitions with ground-based telescopes. The CO ladders of the sample are separated into three classes based on their excitation level.  In 13 of the galaxies, the [OI] 63 $\mu$m emission line is self absorbed. Comparing the CO excitation to the IRAS 60/100 $\mu$m ratio and to far infrared luminosity, we find that the CO excitation is more correlated to the far infrared colors. We present cooling budgets for the galaxies and 
    find fine-structure line flux deficits in the 
    [CII], [SiII], [OI], and [CI] lines in the objects with the highest far IR fluxes, but do not observe this for CO $4\leq J_{upp}\leq13$. In order to study the heating of the molecular gas,
    we present a combination of three diagnostic quantities to help determine the dominant heating source.  Using the CO excitation, the CO J=1-0 linewidth, and the 
    AGN contribution, we conclude that galaxies with large CO linewidths always have 
    high-excitation CO ladders, and often low AGN contributions, 
    suggesting that mechanical heating is important.
   \end{abstract}
 

   \maketitle
%

\section{Introduction}
(Ultra) Luminous Infrared Galaxies ((U)LIRGs) in the local universe are remarkable galaxies exhibiting an extremely high infrared luminosity, 
L$_{8-1000\mu m}>10^{11}$L$_\odot$ for LIRGs and L$_{8-1000\mu m}>10^{12}$L$_\odot$ for ULIRGs. Luminous infrared galaxies were first identified in large numbers
with observations from the InfraRed Astronomical Satellite (IRAS), 
which was launched in 1983 \citep{1985ApJ...290L...5H}. 
After the discovery that these objects 
all contain massive amounts of molecular gas \citep{1988ApJ...325...74S,2002ApJS..143..315V}, detailed studies of the spectroscopic cooling lines were carried out with the 
{\it Infrared Space Observatory} (ISO; \cite{1997ApJ...491L..27M,1998ApJ...504L..11L,2001ApJ...548L..73H,2001ApJ...561..766M,2003ApJ...594..758L,2009ApJ...701.1147A}),
ground based observations of [CI] \citep{1998ApJ...509L..17G,2000ApJ...537..644G}, \emph{Spitzer} Space Telescope \citep{2009PASP..121..559A,2011ApJ...741...32D,2013ApJS..206....1S} and the 
\emph{Herschel} Space Observatory \citep{2011ApJ...728L...7G,2013ApJ...774...68D,2013ApJ...776...38F,2014ApJ...787L..23L,2014ApJ...788L..17D}. In the 
local universe ULIRGs are rare \citep{1991AJ....101..354S}, but at higher redshifts ($z>1$) they represent most of the cosmic infrared background and 
are the dominant source of star formation up to z=2 \citep{2007ApJ...660...97C,2011AA...528A..35M,2011AA...532A..49B,2011A&A...532A..49B,2013A&A...553A.132M,2013MNRAS.436.2875G}. 
Locally, these objects are hosts to intense starbursts, and/or active galactic nuclei (AGN),
and often are part of a merging galaxy group \citep{1987AJ.....94..831A,1988ApJ...328L..35S,1992ARA&A..30..705B,1996ARA&A..34..749S,2002ApJS..143..315V}.  Regardless of the various heating processes available, however, the luminosity of most local (U)LIRGs seem to be
energetically driven by 
starbursts \citep{1998ApJ...498..579G,1998ApJ...507..615D,1999ApJ...522..113V,2002ApJS..143..315V,2004ApJS..152...63G,2009ApJS..182..628V}.  (U)LIRGs are also thought
to represent the transitional phase in evolution from a starburst galaxy to elliptical/lenticular 
galaxies \citep{1988ApJ...328L..35S,2001ApJ...563..527G,2002ApJ...580...73T,2010ApJ...712..318R,2013ApJ...767...72R}, and thus must
quench their star formation during this period. In fact, some evidence for this was found in the discovery of massive molecular outflows with the \emph{Herschel} Space
Observatory \citep{2010A&A...518L..41F,2011ApJ...733L..16S,2013ApJ...775..127S,2013ApJ...776...27V,2014A&A...561A..27G} as well as as by ground-based telescopes (e.g., \citealt{2010A&A...518L.155F}; \citealt{2012ApJ...753..102W}).

Since (U)LIRGs offer a unique insight into this transitional phase from star-forming to quiescent galaxies, understanding which
mechanisms are affecting the star-forming gas is crucial. Many studies of the star forming gas in (U)LIRGs have
been made since its universal presence in (U)LIRGs was determined \citep{1991ApJ...370..158S,1996ARA&A..34..749S,1997ApJ...478..144S}.
In general, gas is heated by either radiation (i.e. UV photons, 
X-ray photons), energetic particles (cosmic rays) or mechanical processes (i.e. turbulence, stellar winds, outflows, supernovae).  The interplay between
these heating sources can account for the extreme environments found in (U)LIRGs, in comparison to less intense
star forming environments \citep{1991AA...249..323A,1995AA...300..369A}.  The high amount
of energy injected into the gas in these galaxies is displayed by emission lines that serve as a coolant along with infrared dust emission. The emission 
lines responsible for most of the gas cooling are the [CII] line at 157 $\mu$m ($^2P_{3/2}-^2P_{1/2}$), the [OI] line at 63 $\mu$m ($^3P_{1}-^3P_{2}$), and CO (rotational transitions). 
The {\it Herschel Space Observatory} has, for the first time, provided astronomers with simultaneous access to these important far infrared cooling lines and the CO rotational ladder (CO ladder) in (U)LIRGs.  Using the multiple rotational 
transitions of CO from J=1-0 through J=13-12, the density, temperature, column density, and mass (with the addition of $^{13}$CO) can be estimated (eg. \citealt{2011ApJ...743...94R,2012ApJ...758..108S,2013MNRAS.434.2051R,2014arXiv1404.6090P}).
In some cases, it is possible to even discern specifically the heating mechanism \citep{2008A&A...488L...5L,2008ApJ...689L.109H,2010AA...518L..42V,2013ApJ...762L..16M,2014ApJ...787L..23L,2014A&A...564A.126R,2014arXiv1404.6470P}.

In this paper, we introduce observations of all major neutral gas cooling line of a representative sample, the HerCULES sample, of local (U)LIRGs spanning the luminosity range 
from 10$^{11} < $L$_{FIR} < 10^{13}$ L$_\odot$. In Section~\ref{sec:obs}, we present the HerCULES sample and observations 
from the {\it Herschel}/SPIRE and {\it Herschel}/PACS spectrometers, which include 
[CII], [OI] 63$\mu$m, [OI] 145$\mu$m, CO ($4\leq J_{upp} <13$), and [CI] 370 $\mu$m and 609 $\mu$m, using the cosmological parameters H$_0$=70 km s$^{-1}$ Mpc, $\Omega_{vaccum}$=0.72, 
and $\Omega_{matter}$=0.28.  In this paper we focus on the main neutral gas cooling lines. 
We therefore do not analyze the [NII] lines, which arise in ionized gas, or other molecular lines which do
not affect the thermal balance. Specifically, we do not discuss H$_2$O since in the cases where these lines are bright, there is strong evidence that they are radiatively excited
\citep{2010A&A...518L..43G,2011ApJ...741L..38V,2013ApJ...771L..24Y} and do not remove kinetic energy from the gas and thus do not contribute to the cooling.  We show spectra for three sample 
galaxies that represent three different classes of excitation, and the CO ladders for the full sample in Section~\ref{sec:results}. Using the full 
sample, in Section~\ref{sec:fir} we analyze the gas excitation, cooling budget of the sample, and a diagram for determining 
additional heating mechanisms for the gas. Our conclusions are presented in Section~\ref{sec:conc}.

\section{Observations}
\label{sec:obs}
\subsection{The HerCULES sample}
The sample was chosen from the IRAS Revised Bright Galaxy Sample (RBGS), 
which contains all 629 extragalactic sources with IRAS 60 $\mu$m flux density S$_{60} > 5.24$ Jy in the (IRAS-covered) sky at Galactic latitudes $|b| > 5$ 
\citep{2003AJ....126.1607S}. From the IRAS RBGS we select a sub-sample applying limits both in S$_{60}$ and $L_{\rm IR}$: at luminosities : $L_{\rm IR} > 10^{12}$ L$_\odot$ (ULIRGs), 
all sources with S$_{60} > 11.65$ Jy are included, while at luminosities
10$^{11}$ L$_\odot < L_{\rm IR} < 10^{12}$ L$_\odot$ (LIRGs), sources with S$_{60} > 16.4$ Jy are included. From this flux-limited representative parent sample of 32 targets, 
we removed three LIRGs for which no ground-based CO data are available, with the exception of ESO 173-G015, IRAS 13120-5453, and MCG+12-02-001. The resulting representative flux-limited sample consists of 21 LIRGs and 8 ULIRGs. The sample covers a 
factor of 32 in $L_{\rm IR}$ and contains a range of objects including starburst galaxies, AGNs, and composite sources, and covering also a range of IRAS 60/100 micron ratios. 
The full list of included galaxies and their respective properties can be found in Table~\ref{tab:obs}. The infrared 
luminosity and the luminosity distance are from \citet{2009PASP..121..559A}.

In order to obtain a comprehensive view of the CO emission and the cooling budget of these galaxies, we proposed \emph{Herschel}/SPIRE spectroscopy (for the CO ladder) and \emph{Herschel}/PACS 
spectroscopy (for the [CII] and [OI] fine structure lines) of the entire sample, unless PACS observations were already observed as part of another program. In addition to the galaxies observed for HerCULES, we have included NGC 4418, NGC 1068, and Arp 220 for completeness.
This project was approved as a Key Project on the Herschel Space Observatory, under the name 
Herschel Comprehensive (U)LIRG Emission Survey (HerCULES - P.I. Van der Werf). Key elements of HerCULES are:
\begin{itemize}
 \item a representative flux-limited sample of local LIRGs and ULIRGs;
  \item comprehensive coverage of the SPIRE spectral range at the highest spectral resolution mode (covering the CO ladder, [CI] and [NII] fine structure lines, and any other bright features such as H$_2$O lines);
 \item comprehensive coverage of the key fine-structure cooling lines [CII] and [OI] with PACS observations;
\end{itemize}
Details about the galaxy type and observation ID can be seen in Table~\ref{tab:obs}. We have included observations from other programs (KPGT\_esturm\_1K, KPGT\_cwilso01\_1, OT1\_larmus\_1,
OT1\_shaileyd\_1) to help realize the complete flux-limited sample.  The references
for these observations are also in Table~\ref{tab:obs}.

\subsection{Herschel/SPIRE observations}
Spectra were obtained with the Spectral and Photometric Imaging Receiver and
Fourier-Transform Spectrometer \citep[SPIRE-FTS,][]{2010AA...518L...3G} on board
the \emph{Herschel Space Observatory} \citep{2010AA...518L...1P} for the full HerCULES sample.  The observations were carried out in staring mode with the galaxy nucleus on the central pixel of the detector array, with 
a beam size varying from 17''-42'' for the CO transitions.  The high spectral resolution mode was used with
a resolution of 1.2 GHz over the two observing bands.  The low frequency
focal plane array (Long Wavelength Spectrometer Array, SLW) covers $\nu$=447-989 GHz ($\lambda$=671-303 $\mu$m) and the high
frequency focal plane array (Short Wavelength Spectrometer Array, SSW) covers $\nu$=958-1545 GHz ($\lambda$=313-194 $\mu$m),
and together they include the CO $J=4-3$ to CO $J=13-12$ lines.  For galaxies with $z>0.03$, the rest frequency of the J=4-3 transition falls short of the SPIRE coverage. All 
galaxies were observed in the sparse observing mode besides NGC 4418, which was observed in the intermediate mode. 

The data were reduced using version 13.0 of the Herschel Interactive Processing
Environment (HIPE). Initial processing steps included timeline deglitching, linearity correction, clipping of saturated points, time-domain phase correction, 
and inteferogram baseline subtraction. After a second deglitching step and interferogram phase correction, the interferograms 
were Fourier transformed, and the thermal emission from instrument and telescope was removed from the resulting spectra. The 
averaged  spectra were flux calibrated as point sources using the calibration tree associated with HIPE 13.0.
Following these steps a "dark" spectrum was subtracted, to remove any residual emission from the telescope and the instrument. 
Since the emission of most of our sources is contained entirely in the central pixel of the detector arrays, a "dark" spectrum 
was constructed by spectrally smoothing a combination of several off-axis pixels. For extended targets, where the off-axis pixels 
contain emission, the dark was obtained from a deep blank-sky observation obtained on the same observing day.  We compared the two methods and found no differences, but the noise was smaller using the smooth off axis pixels, in the case of the compact targets.

For all extended sources (Arp 299, ESO 173-G015,MCG+12--02--001, Mrk 331, NGC 1068, NGC 1365, NGC 2146, NGC 3256, NGC 5135, and NGC 7771), an aperture correction is
necessary to compensate for the wavelength dependent beam size \citep{2013ApOpt..52.3864M}. We defined a source as extended using LABOCA or SCUBA 350 or 450 $\mu$m (respectively)
maps with 8'' resolution.  We convolved the 8'' resolution maps with the SPIRE FTS resolution, and if the galaxy was more extended than the smallest SPIRE beam size, 
we defined it as extended. In order to correct for the extended nature of these sources, we
employ HIPE's semiExtendedCorrector tool (SECT).  This tool 'derives' an intrinsic source size by
iterating over different source sizes until it finds one that provides
a good match in the overlap range of the two observing bands near 1000 GHz, and is further discussed in \citet{2013A&A...556A.116W}. We set the Gaussian reference beam 
to 42'', the largest SPIRE beamsize. The beamsize corrected flux values for the 10 extended sources are listed in Table~\ref{tab:flux}, along with the 
compact sources. We note that the error in the extended source flux correction could be significant due to the assumptions that the
high-J CO transitions are distributed in the same way as the the low-J CO lines.  If high-J CO transitions are only coming from a centralized
compact region, we are overestimating their flux with our beam correction method.  For this reason, we apply an additional 15\% error
to the extended galaxies. There 
are three targets in the sample that have multiple pointings; Arp 299, NGC 1365, and NGC 2146.  In the case of Arp 299, we use only the pointing 
for Arp 299 A.  For NGC 1365 we take the average of the northeast and southwest pointings. This is done since the northeast and southwest 
pointings have approximately a 50\% overlap in field of view at the center of the galaxy.  This overlap region is the center of the 
PACS observations, so for comparison, it is best to average the northeast and southwest pointings. For NGC 2146, we use only the nuclear pointing.

CO and [CI] line fluxes were extracted using version 1.92 of FTFitter\footnote{https://www.uleth.ca/phy/naylor/index.php?page=ftfitter}, 
a program specifically created to extract line fluxes from Fourier transform spectrographs, and are listed in Table~\ref{tab:flux}.  This is an interactive 
data language (IDL) based graphical user interface, that allows the user to fit lines, choose line 
profiles, fix any line parameter, and extract the flux.
We define a third order polynomial baseline to fit the continuum for the SLW and SSW separately and derive the integrated line intensities from baseline subtracted spectra with a simultaneous line fit of all
CO, [CI], [NII] and other bright lines in the spectrum. We use a Gaussian line profile, which is based on the assumed intrinsic line shape with a width derived from spectrally resolved CO 1-0, convolved
with the instrumental line shape, which is a Sinc profile.
We adopt an error of 16\% for the non-extended galaxy fluxes, which encompasses
our dominant sources of error, 10\% for the flux extraction and baseline definition and 6\% for the absolute
calibration uncertainty for staring-mode SPIRE FTS
observations \citep{2014arXiv1403.1107S}.  For the case of extended sources, we adopt as already mentioned, an additional 15\% error from the beam size
corrections, resulting in a total error 30\% for the 10 extended sources.

\begin{deluxetable}{lcccccccccc}
\setlength{\tabcolsep}{3pt}
\tabletypesize{\scriptsize}
\tablecaption{Sample Properties\label{tab:obs}}
\tablewidth{\textwidth}
\tablehead{\colhead{1} & \colhead{2} & \colhead{3} & \colhead{4} & \colhead{5}&\colhead{6} & \colhead{7} &\colhead{8}& \colhead{9}&\colhead{10} & \colhead{11} }
\startdata
NGC 34			&11.49	&0.78	&0.01962 & 84.1		&330	& SB			&[OI]$_{63}$, [OI]$_{145}$,[CII]&00$^h$11$^m$06.67$^s$ -12\degree06'26.13''	&1342199416&KPOT\_pvanderw\_1\\
(IRAS 00085-1223) 	&	&	&	&		&	&			&$194-671$~$\mu$m		&00$^h$11$^m$06.53$^s$ -12\degree06'24.90''	&1342199253&KPOT\_pvanderw\_1\\
			
MCG+12--02--001		&11.50	&1.07	&0.01570 & 69.8		&200	& SB			&[OI]$_{63}$, [OI]$_{145}$,[CII]&00$^h$54$^m$03.33$^s$ +73\degree04'59.83''	&1342193211&KPOT\_pvanderw\_1\\
(IRAS 00506+7248) 	&	&	&	&		&	&			&$194-671$~$\mu$m		&00$^h$54$^m$03.56$^s$ +73\degree05'10.38''	&1342213377&KPOT\_pvanderw\_1\\
			
IC 1623			&11.71	&1.14	&0.02007 & 85.5 	&250	&SB,AGN			&[OI]$_{63}$, [OI]$_{145}$,[CII]&01$^h$07$^m$46.59$^s$ -17\degree30'26.46''	&1342212532&KPOT\_pvanderw\_1\\	
(IRAS 01053-1746)		&	&	&	 &		&	&			&$194-671$~$\mu$m		&01$^h$07$^m$46.74$^s$ -17\degree30'26.05''	&1342212314&KPOT\_pvanderw\_1\\

NGC 1068		&11.40	&9.07	&0.003793& 15.9		&300	& AGN,SB		&[OI]$_{63}$			&02$^h$42$^m$40.78$^s$ -00\degree00'47.16''	&1342191153&KPGT\_esturm\_1K\\
(IRAS 02401-0013) 	&	&	&	&		&	&			&[OI]$_{145}$,[CII]		&02$^h$42$^m$40.73$^s$ -00\degree00'42.24''	&1342191154&KPGT\_esturm\_1K \\
			&	&	&	&		&	&			&$194-671$~$\mu$m		&02$^h$42$^m$40.92$^s$ -00\degree00'46.65''	&1342213445&KPGT\_cwilso01\_1\\
NGC 1365		&11.00	&4.32	&0.00546 & 17.9		&250	& Sy1,SB		&[OI]$_{63}$			&03$^h$33$^m$36.31$^s$ -36\degree08'16.61''	&1342191295&KPGT\_esturm\_1K\\
(IRAS 03317-3618) 	&	&	&	&		&	&			&[OI]$_{145}$,[CII]		&03$^h$33$^m$36.26$^s$ -36\degree08'24.33''	&1342191294&KPGT\_esturm\_1K\\
			&	&	&	&		&	&			&$194-671$~$\mu$m		&03$^h$33$^m$36.48$^s$ -36\degree08'19.32''	&1342204020&KPOT\_pvanderw\_1\\
NGC 1614		&11.65	&1.50	&0.01594 & 67.8 	&220	& SB			&[OI]$_{63}$, [OI]$_{145}$,[CII]&04$^h$33$^m$59.79$^s$ -08\degree34'44.19''	&1342190367&KPOT\_pvanderw\_1\\
(IRAS 04315-0840) 	&	&	&	&		&	&			&$194-671$~$\mu$m		&04$^h$33$^m$59.85$^s$ -08\degree34'44.15''	&1342192831&KPOT\_pvanderw\_1\\
				
IRAS F05189--2524	&12.16	&0.60	&0.04256 & 187		&300	& QSO			&[OI]$_{63}$			&05$^h$21$^m$01.24$^s$ -25\degree21'43.16''	&1342219441&KPGT\_esturm\_1K\\
			&	&	&	&		&	&			&[OI]$_{145}$,[CII]		&05$^h$21$^m$01.28$^s$ -25\degree21'42.15''	&1342219442&KPGT\_esturm\_1K\\
			&	&	&	&		&	&			&$194-671$~$\mu$m		&05$^h$21$^m$01.42$^s$ -25\degree21'45.47''	&1342192832\tablenotemark{a}&KPOT\_pvanderw\_1\\
			&	&	&	&		&	&			&$194-671$~$\mu$m		&05$^h$21$^m$01.42$^s$ -25\degree21'45.48''	&1342192833\tablenotemark{a}&KPOT\_pvanderw\_1\\
NGC 2146		&11.12	&6.97	&0.00298 & 17.5		&250	& SB			&[OI]$_{63}$, [OI]$_{145}$,[CII]&06$^h$18$^m$35.53$^s$ +78\degree21'25.39''	&1342193210&KPOT\_pvanderw\_1\\
(IRAS 06106+7822) 	&	&	&	&		&	&			&$194-671$~$\mu$m		&06$^h$18$^m$38.07$^s$ +78\degree21'25.06''	&1342204025&KPOT\_pvanderw\_1\\
			
NGC 2623 		&11.60	&1.15	&0.01851 & 84.1		&400	& SB,AGN		&[OI]$_{63}$, [OI]$_{145}$,[CII]&08$^h$38$^m$24.29$^s$ +25\degree45'16.72''	&1342208904&KPOT\_pvanderw\_1\\
(IRAS 08354+2555) 	&	&	&	&		&	&			&$194-671$~$\mu$m		&08$^h$38$^m$24.14$^s$ +25\degree45'17.34''	&1342219553&KPOT\_pvanderw\_1\\
			
NGC 3256		&11.64	&4.61	&0.00935 & 38.9		&230	& SB			&[OI]$_{63}$			&10$^h$27$^m$51.61$^s$ -43\degree54'15.39''	&1342210383&KPGT\_esturm\_1K\\
(IRAS 10257-4338) 	&	&	&	&		&	&			&[OI]$_{145}$,[CII]		&10$^h$27$^m$51.45$^s$ -43\degree54'21.87''	&1342210384&KPGT\_esturm\_1K\\
			&	&	&	&		&	&			&$194-671$~$\mu$m		&10$^h$27$^m$51.49$^s$ -43\degree54'16.00''	&1342201201&KPOT\_pvanderw\_1\\
Arp 299 A 		&11.93 	&4.84	&0.01030 & 50.7 	&325 	& SB,AGN		&[OI]$_{63}$ 			&11$^h$28$^m$33.41$^s$ +58\degree33'46.04''	&1342199421 &KPGT\_esturm\_1K \\
IC 694			&	&	&	&		&	&			&[OI]$_{145}$			&11$^h$28$^m$33.41$^s$ +58\degree33'46.04''	&1342232602&OT1\_shaileyd\_1\\
(IRAS 11257+5850) 	&	&	&	&		&	&			&[CII]				&11$^h$28$^m$33.41$^s$ +58\degree33'46.04''	&1342208906&KPGT\_esturm\_1K \\
			&	&	&	&		&	&			&$194-671$~$\mu$m		&11$^h$28$^m$33.41$^s$ +58\degree33'46.04''	& 1342199248&KPOT\_pvanderw\_1\\
ESO 320--G030 		&11.17  &1.67	&0.01078 & 41.2 	&350 	& SB 			&[OI]$_{63}$, [OI]$_{145}$,[CII]&11$^h$53$^m$11.75$^s$ -39\degree07'51.75''	&1342212227&KPOT\_pvanderw\_1\\
(IRAS11506--3851)		&	&	&	 &		&	&			&$194-671$~$\mu$m		&11$^h$53$^m$11.52$^s$ -39\degree07'50.24''	&1342210861&KPOT\_pvanderw\_1\\	

NGC 4418		&11.19	&1.74	&0.007268 & 36.5	&163	& Sy2			&[OI]$_{63}$			&12$^h$26$^m$54.51$^s$ -00\degree52'40.77''	&1342187780&KPGT\_esturm\_1K\\
(IRAS 12243-0036) 	&	&	&	&		&	&			&[OI]$_{145}$,[CII]		&12$^h$26$^m$54.57$^s$ -00\degree52'36.93''	&1342210830&KPGT\_esturm\_1K\\
			&	&	&	&		&	&			&$194-671$~$\mu$m		&12$^h$26$^m$54.60$^s$ -00\degree52'36.54''	&1342210848&KPGT\_esturm\_1K\\
Mrk 231			&12.57	&1.48	&0.04217 & 192 		&200	& QSO			&[OI]$_{63}$			&12$^h$56$^m$14.65$^s$ +56\degree52'24.13''	&1342189280&KPGT\_esturm\_1K\\
(IRAS 12540+5708) 	&	&	&	&		&	&			& [OI]$_{145}$,[CII]		&12$^h$56$^m$14.29$^s$ +56\degree52'23.40''	&1342186811&SDP\_esturm\_3\\
			&	&	&	&		&	&			&$194-671$~$\mu$m		&12$^h$56$^m$14.29$^s$ +56\degree52'26.73''	&1342210493&KPOT\_pvanderw\_1\\
IRAS13120--5453		&12.32	&1.94	&0.03076 & 144 		&400 	& Sy2,SB		&[OI]$_{63}$			&13$^h$15$^m$06.28$^s$ -55\degree09'24.46''	&1342214628&KPGT\_esturm\_1K\\
			&	&	&	&		&	&			&[OI]$_{145}$,[CII]		&13$^h$15$^m$06.17$^s$ -55\degree09'25.38''	&1342214629&KPGT\_esturm\_1K\\
			&	&	&	&		&	&			&$194-671$~$\mu$m		&13$^h$15$^m$06.11$^s$ -55\degree09'23.21''	&1342212342&KPOT\_pvanderw\_1\\
Arp 193 		&11.73	&8.19	&0.02330 & 110		&400 	& SB,L 			&[OI]$_{63}$, [OI]$_{145}$,[CII]&13$^h$20$^m$35.20$^s$ +34\degree08'24.58''  	&1342197801	&KPOT\_pvanderw\_1 \\
(IRAS 13183+3423)		&	&	&	 &		&	&			&$194-671\mu$m			&13$^h$20$^m$35.35$^s$ +34\degree08'23.46'' 	&1342209853&KPOT\_pvanderw\_1\\
NGC 5135		&11.30	&0.91	&0.01369 & 60.9		&150	& Sy2,SB		&[OI]$_{63}$, [OI]$_{145}$,[CII]&13$^h$25$^m$43.96$^s$ -29\degree50'01.74''	&1342190371&KPOT\_pvanderw\_1\\
(IRAS 13229-2934)		&	&	&	&		&	&			&$194-671$~$\mu$m		&13$^h$25$^m$43.91$^s$ -29\degree50'00.27'' 	&1342212344&KPOT\_pvanderw\_1\\
ESO 173--G015		&11.38 	&3.61	&0.00974& 34 		&200	& SB 			&[OI]$_{63}$,[OI]$_{145}$,[CII]	&13$^h$27$^m$24.00$^s$ -57\degree29'23.63''	&1342190368&KPOT\_pvanderw\_1\\
(IRAS13242--5713)		&	&	&	&		&	&			&$194-671$~$\mu$m		&13$^h$27$^m$23.95$^s$ -57\degree29'22.89'' 	&1342202268&KPOT\_pvanderw\_1\\	
Mrk 273			&12.21	&1.05	&0.03736 & 173		&520	& SB,Sy2		&[OI]$_{63}$			&13$^h$44$^m$42.09$^s$ +55\degree53'09.14''	&1342207801&KPGT\_esturm\_1K\\
(IRAS 13428+5608) 	&	&	&	&		&	&			&[OI]$_{145}$,[CII]		&13$^h$44$^m$41.82$^s$ +55\degree53'08.75''	&1342207802&KPGT\_esturm\_1K \\
			&	&	&	&		&	&			&$194-671$~$\mu$m		&13$^h$44$^m$42.10$^s$ +55\degree53'10.50''	&1342209850&KPOT\_pvanderw\_1\\
Zw 049.057		&11.35	&1.05	&0.01300&65.4		&200	&SB			&[OI]$_{63}$, [OI]$_{145}$,[CII]&15$^h$13$^m$13.18$^s$ +07\degree13'30.71''	&1342190374&KPOT\_pvanderw\_1\\
CGCG 049-057		&	&	&	&		&	&			&$194-671$~$\mu$m		&15$^h$13$^m$13.10$^s$ +07\degree13'29.19''	&1342212346&KPOT\_pvanderw\_1\\
(IRAS 15107+0724) 	&	&	&	&		&	&			&				&						&		&		\\

Arp 220 		&12.28	&4.87	&0.01813 & 77		&504	& SB,AGN		&[OI]$_{63}$, [OI]$_{145}$	&15$^h$34$^m$57.22$^s$ +23\degree30'11.06''	&1342191304&KPGT\_esturm\_1K \\
(IRAS 15327+2340) 	&	&	&	&		&	&			&[CII]				&15$^h$34$^m$57.21$^s$ +23\degree30'10.13''	&1342191306&KPGT\_esturm\_1K \\
			&	&	&	&		&	&			&$194-671$~$\mu$m		&15$^h$34$^m$57.11$^s$ +23\degree30'11.26'' 	&1342190674&KPGT\_cwilso01\_1 \\
NGC 6240		&11.93	&1.10	&0.02448 & 116		&500	& SB,AGN		&[OI]$_{63}$			&16$^h$52$^m$59.10$^s$ +02\degree24'03.58''	&1342216622&KPGT\_esturm\_1K\\
(IRAS 16504+0228) 	&	&	&	&		&	&			&[OI]$_{145}$,[CII]		&16$^h$52$^m$59.10$^s$ +02\degree24'02.79''	&1342216623&KPGT\_esturm\_1K\\
			&	&	&	&		&	&			&$194-671$~$\mu$m		&16$^h$52$^m$59.01$^s$ +02\degree24'03.27''	&1342214831&KPOT\_pvanderw\_1\\
IRAS F17207--0014	&12.46	&1.56	&0.04281 & 198		&620	& SB,L			&[OI]$_{63}$			&17$^h$23$^m$21.83$^s$ -00\degree16'59.82''	&1342229692&KPGT\_esturm\_1K\\
			&	&	&	&		&	&			&[OI]$_{145}$,[CII]		&17$^h$23$^m$31.84$^s$ -00\degree16'57.60''	&1342229693& KPGT\_esturm\_1K\\	
			&	&	&	&		&	&			&$194-671$~$\mu$m		&17$^h$23$^m$21.93$^s$ -00\degree17'01.10''	&1342192829&KPOT\_pvanderw\_1\\
IRAS F18293--3413	&11.88	&1.82	&0.01817 & 86		&270	& 			&[OI]$_{63}$, [OI]$_{145}$,[CII]&18$^h$32$^m$41.34$^s$ -34\degree11'36.90''	&1342192112&KPOT\_pvanderw\_\\
			&	&	&	&		&	&			&$194-671$~$\mu$m		&18$^h$32$^m$41.17$^s$ -34\degree11'27.23''	&1342192830&KPOT\_pvanderw\_1\\

IC 4687/6 		&11.62	&0.84	&0.01735 & 81.9 	&230	& SB			&[OI]$_{63}$			&18$^h$13$^m$39.94$^s$ -57\degree43'49.66''	&1342239740&OT1\_larmus\_1\\
(IRAS 18093-5744) 	&	&	&	&		&	&			&[CII]				&18$^h$13$^m$39.80$^s$ -57\degree43'35.71''	&1342239739&OT1\_larmus\_1 \\
			&	&	&	&		&	&			&$194-671$~$\mu$m		&18$^h$13$^m$39.50$^s$ -57\degree43'31.05''	&1342192993&KPOT\_pvanderw\_1\\

NGC 7469		&11.65	&1.32	&0.01632 & 70.8		&300	& Sy1,SB		&[OI]$_{63}$			&23$^h$03$^m$15.47$^s$ +08\degree52'37.05''	&1342187847&KPGT\_esturm\_1K\\
(IRAS 23007+0836)		&	&	&	&		&	&			&[OI]$_{145}$,[CII]		&23$^h$03$^m$15.83$^s$ +08\degree52'28.52''	&1342211171&KPGT\_esturm\_1K\\
			&	&	&	&		&	&			&$194-671$~$\mu$m		&23$^h$03$^m$15.87$^s$ +08\degree52'28.15''	&1342199252&KPOT\_pvanderw\_1\\
			
NGC 7552		&11.11	&3.64	&0.00537 & 23.5		&180	& SB			&[OI]$_{63}$			&23$^h$16$^m$10.10$^s$ -42\degree34'53.89''	&1342210400&KPGT\_esturm\_1K\\
(IRAS 23134-4251) 	&	&	&	&		&	&			&[OI]$_{145}$,[CII]		&23$^h$16$^m$10.59$^s$ -42\degree35'05.73''	&1342210399&KPGT\_esturm\_1K \\
			&	&	&	&		&	&			&$194-671$~$\mu$m		&23$^h$16$^m$10.73$^s$ -42\degree35'06.02''	&1342198428&KPOT\_pvanderw\_1\\
			
NGC 7771		&11.40	&1.08	&0.01427 & 61.2		&250	& SB			&[OI]$_{63}$, [OI]$_{145}$,[CII]&23$^h$51$^m$24.83$^s$ +20\degree06'42.33''	&1342197839&KPOT\_pvanderw\_1\\
(IRAS 23488+1949) 	&	&	&	&		&	&			&$194-671$~$\mu$m		&23$^h$51$^m$24.72$^s$ +20\degree06'42.11''	&1342212317&KPOT\_pvanderw\_1\\

Mrk 331 		&11.50	&0.87	&0.01790 & 79.3		&215	& SB			&[OI]$_{63}$, [OI]$_{145}$,[CII]&23$^h$51$^m$26.76$^s$ +20\degree35'09.83''	&1342197840&KPOT\_pvanderw\_1\\
(IRAS 23488+2018) 	&	&	&	&		&	&			&$194-671$~$\mu$m		&23$^h$51$^m$26.65$^s$ +20\degree35'10.42''	&1342212316&KPOT\_pvanderw\_1\\

\enddata
 \tablecomments{\\
Column 1: Object name. \\
Column 2: log($L_{\rm IR}$/L$_\odot$) from \citet{2009PASP..121..559A}. Observations use the cosmological parameters H$_0$=70 km s$^{-1}$ Mpc, $\Omega_{vaccum}$=0.72, 
 and $\Omega_{matter}$=0.28.\\
Column 3: Far infrared flux (FIR) calculated using the IRAS definition \citep{1985ApJ...298L...7H} in 10$^{-12}$ W m$^{-2}$. \\
Column 4: Redshift $z$ from NED.\\
Column 5: Luminosity distance D$_L$ in Mpc from \citet{2009PASP..121..559A}.\\
Column 6: CO 1-0 full width to half power line width in km s$^{-1}$. \\
Column 7: Galaxy classification from NED SB=Starburst, L=LINER, AGN=Active Galaxy Nucleus, Sy1=Seyfert 1, Sy2=Seyfert 2, QSO=Quasi-Stellar Object. \\
Column 8: Line names. \\
Column 9: Pointing Coordinates.\\
Column 10: Observation ID (OBSID). \\
Column 11: Program ID. \\}
 
\tablenotetext{a}{The two SPIRE observations of IRAS F05189--2524 were combined using HIPE average (avg) task.} 
\end{deluxetable}

\subsection{Herschel/PACS observations}
We have obtained observations of the [OI] 63 $\mu$m ([OI]$_{63}$), [OI] 145 $\mu$m ([OI]$_{145}$), and [CII] 157 $\mu$m emission lines with the
Integral Field Spectrometer of the Photodetector Array Camera and Spectrometer
\citep[PACS,][]{2010AA...518L...2P} on board {\it Herschel Space Observatory} for every object in the HerCULES sample. The data presented
here were obtained as part of the Herschel program KPOT\_pvanderw\_1
(PI: P. van der Werf), complemented by observations from other programs. The observations and program IDs of the [CII] and [OI] lines are listed in
Table~\ref{tab:obs}.

The data were downloaded from the Herschel Science Archive and processed
using HIPE v11.0.  Standard processing steps including timeline deglitching, application of the Relative Spectral Response 
Function and detector flat fielding, and subtraction of the on and off chop positions, gridding along the spectral axis, and combination 
of the nod positions.  With the exception of Arp 299, the objects
are all centered on the 9$\,.\!\!^{\prime\prime}$4 central spaxel of the 5 by 5 PACS
array, observed in staring mode. 
The fluxes are extracted from the central spaxel, using the extractSpaxelSpectrum routine, and referenced to a
point source. We use the pointSourceLossCorrection routine to capture any 
additional flux that may not be captured in the central spaxel. Finally, version 3.10 of SPLAT as part of the STARLINK software package (http://star-www.dur.ac.uk/~pdraper/splat/splat-vo/) was used to
subtract the baseline from each observation, and isolate the desired lines, in the case of PACS range spectroscopy.  The reduction steps were the same for both the PACS range and line spectroscopy, two different
observing modes of PACS.

Arp 299 was observed in the mapping mode and reduced using the standard pipeline reduction.  The integrated flux for Arp 299 A (presented in this paper) was 
calculated by summing the flux within a 25'' aperture centered on Arp 299 A SPIRE pointing.

In order to extract the line parameters from the PACS observations, we first integrate over the baseline subtracted spectrum and then we 
fit a Gaussian profile to the baseline subtracted flux. In some sources, the [OI]$_{63}$ 
line shows a double-peaked profile, where the flux at the central wavelength is diminished, which could indicate Keplerian rotation.  
However, if this were the case, then we would expect a similar profile in the [OI] 145 $\mu$m line and possibly the other fine structure lines as well, which is not seen. The spectral resolution of 
PACS
at 145 $\mu$m is more than sufficient to resolve the $\sim0.2$ $\mu$m separation between the two peaks in the [OI]$_{63}$ profile (Figure~\ref{fig:specn}). Therefore, we conclude that this double-peaked profile is 
due to absorption in the center of the profile by colder foreground gas.  We note that [OI]$_{63}$ absorption is due to O in the ground state while absorption at 145$\mu$m requires O to be at a state having 
an energy of 226 K above the ground state.  Therefore, in cool or moderate density gas the [OI]$_{145}$ line will not show an absorption feature, even if the [OI]$_{63}$ line does.  This same effect has been noted in 
Arp 220, which shows the [OI]$_{63}$ in full absorption \citep{2012A&A...541A...4G}.  In the case of
NGC 4418 and Zw 049.057, the [OI]$_{63}$ line has an inverse P Cygni profile, suggesting that the absorbing foreground gas is flowing into the nuclear region.  The three example galaxies for which 
the spectra are shown in Figures 1-3 display increasing absorption of the [OI]$_{63}$ $\mu$m line.  In NGC 7552, a face-on 
starburst galaxy, the profile remains Gaussian, while in Mrk 331, a late-stage merger, there is a strong dip in the middle of the profile.  IRAS F17207--0014 is known for being one of the coolest 
ULIRGs, here absorption dominates the [OI]$_{63}$ emission.  For the [OI]$_{63}$ profiles that show an absorption feature, we fit the Gaussian only 
to the wings of the emission profile and state the flux in parentheses. The Gaussian-fit flux is only valid if the true line profile is Gaussian.  We suggest this is a more robust estimate of the true integrated flux of the [OI]$_{63}$ line emerging from the warm nuclear region, since in many cases, the absorption dominates
the profile. We note that using a Gaussian profile to extrapolate the line flux requires an assumption that the location of the line center (a free parameter) is in the middle of the profile, which may not be accurate, especially in the case of 
IRAS F17207--0014, or any other galaxies with an asymmetric line profile.  We have tested the relations presented in the rest of this paper with both the integrated flux and the Gaussian fit, and find it does not strongly affect the results. Both the observed line fluxes and the 
gaussian-fit line fluxes, stated in the parentheses, are also presented in Table~\ref{tab:flux}. 

\section{Results}
\label{sec:results}
\subsection{Spectra and line fluxes}
\begin{figure*}[ht]
\epsscale{2.}
\plotone{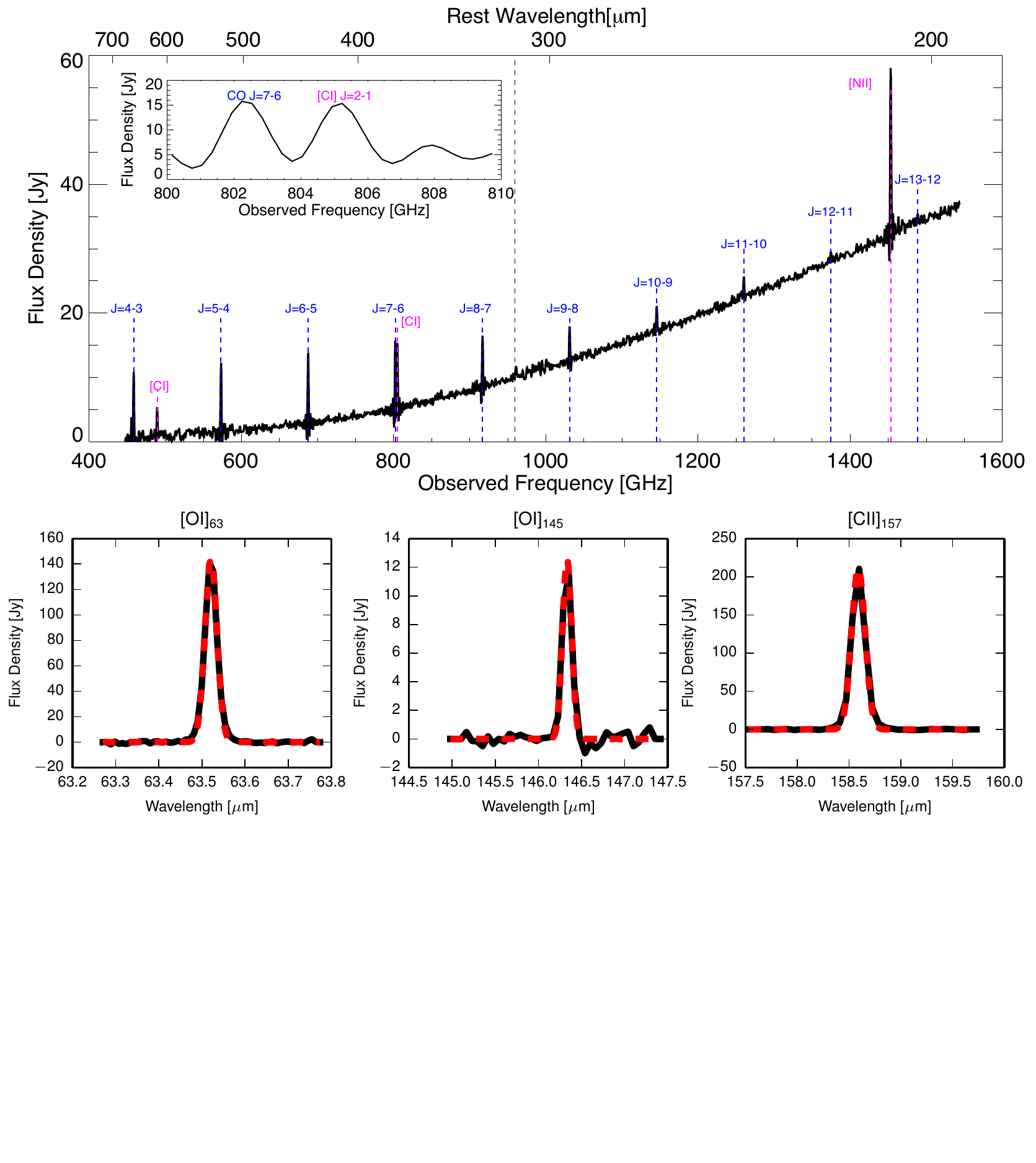}
\caption{Top: Herschel SPIRE spectrum for NGC 7552 in observed frequency.  CO lines are indicated in blue and fine structure lines in pink. The vertical black dashed line near 920 GHz separates the SLW and SSW arrays.  
The inset shows a magnified version of the CO J=7-6 and 
[CI]$_{370}$ transitions. Bottom: Baseline subtracted Herschel PACS observations of [OI]$_{63}$, [OI]$_{145}$, and [CII] presented in black, with superimposed Gaussian line fits shown in red.\label{fig:specn}}
\end{figure*}

\begin{figure*}[ht]
\epsscale{2.}
\plotone{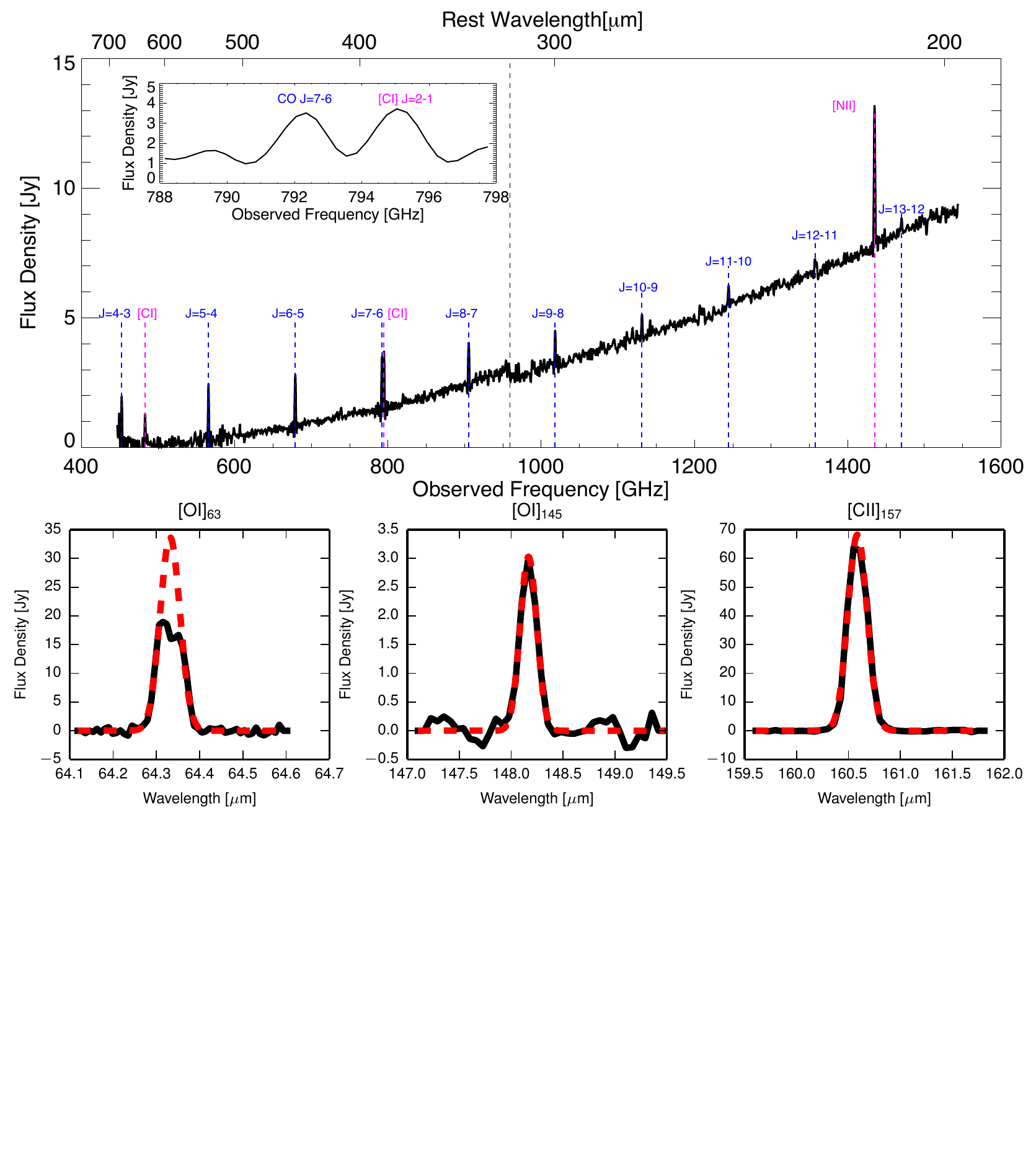}
\caption{Top: Herschel SPIRE spectrum for Mrk 331 in observed frequency.  CO lines are indicated in blue and fine structure lines in pink. The vertical black dashed line near 920 GHz separates the SLW and SSW arrays.  
The inset shows a magnified version of the CO J=7-6 and 
[CI]$_{370}$ transitions. Bottom: Baseline subtracted Herschel PACS observations of [OI]$_{63}$, [OI]$_{145}$, and [CII] presented in black, with superimposed Gaussian line fits shown in red.}
\label{fig:speca}
\end{figure*}

\begin{figure*}[ht]
\epsscale{2.}
\plotone{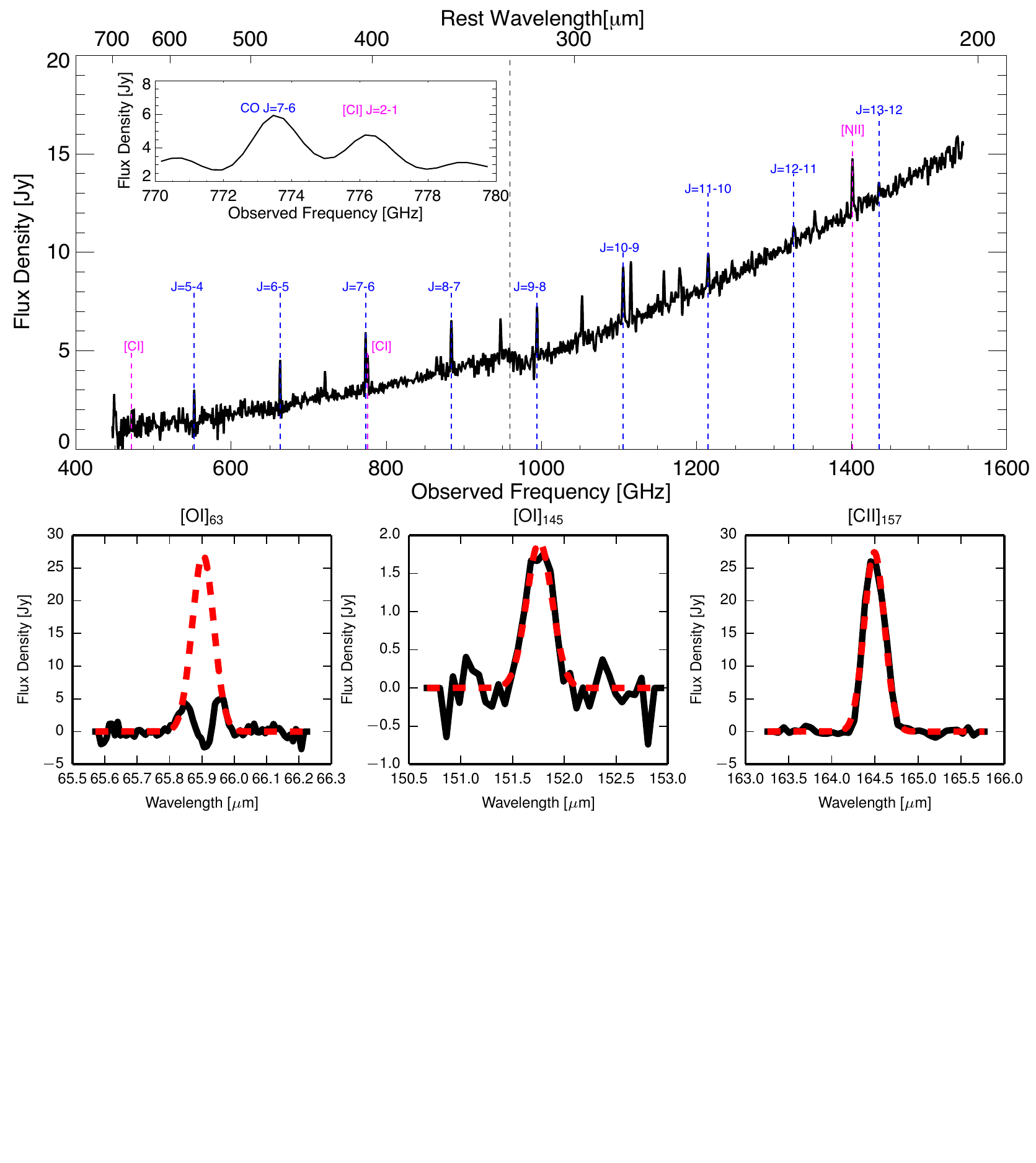}
\caption{Top: Herschel SPIRE spectrum for IRAS F17207--0014 in observed frequency.  CO lines are indicated in blue and fine structure lines in pink. The vertical black dashed line near 920 GHz separates the SLW and SSW arrays.  
The inset shows a magnified version of the CO J=7-6 and 
[CI]$_{370}$ transitions. Bottom: Baseline subtracted Herschel PACS observations of [OI]$_{63}$, [OI]$_{145}$, and [CII] presented in black, with superimposed Gaussian line fits shown in red.}
\label{fig:speci}
\end{figure*}

All SPIRE CO and [CI] line fluxes are 
listed in Table~\ref{tab:flux}.  We present three examples of galaxy spectra obtained with SPIRE in the top panels of Figure~\ref{fig:specn},~\ref{fig:speca}, and ~\ref{fig:speci} for
NGC 7552, Mrk 331, and IRAS F17207--0014, respectively.  It is important to note 
that the baseline ripple seen in the SPIRE FTS spectra is due to both the sinc profile of the strong CO transitions and the noise. Since in this paper we 
only discuss the neutral gas cooling, we do not present fluxes of [NII] (which originates in ionized gas) or the
molecular lines other than CO, which are irrelevant to the total neutral gas cooling. A comprehensive set of fluxes will be presented in Van der Werf et al. (in prep).  In addition, the HerMES team is planning to publish a formal data paper using HIPE v12.0 with a detailed error analysis in the near future.

In the bottom row of the spectra in Figures~\ref{fig:specn},~\ref{fig:speca}, and~\ref{fig:speci}, the PACS line profiles of the three sample galaxies are 
presented (NGC 7552, Mrk 331, IRAS F17207--0014).  

\begin{sidewaystable*}
\resizebox{\columnwidth}{!}{
\begin{tabular}{lccccccccccccccc}
 
\hline
\hline
Object 			& CO J=4-3& CO J=5-4& CO J=6-5& CO J=7-6& CO J=8-7& CO J=9-8& CO J=10-9& CO J=11-10& CO J=12-11& CO J=13-12&[CI]$_{609}$&[CI]$_{370}$&[OI]$_{63}$&[OI]$_{145}$ &[CII] \\
& 10$^{-17}$ Wm$^{-2}$& 10$^{-17}$ Wm$^{-2}$& 10$^{-17}$ Wm$^{-2}$& 10$^{-17}$ Wm$^{-2}$& 10$^{-17}$ Wm$^{-2}$& 10$^{-17}$ Wm$^{-2}$&10$^{-17}$ Wm$^{-2}$&10$^{-17}$ Wm$^{-2}$& 10$^{-17}$ Wm$^{-2}$&10$^{-17}$ Wm$^{-2}$&10$^{-17}$ Wm$^{-2}$&10$^{-17}$ Wm$^{-2}$&10$^{-15}$ Wm$^{-2}$&10$^{-15}$ Wm$^{-2}$ &10$^{-15}$ Wm$^{-2}$ \\
\hline
NGC 34			&--&      1.66&      2.23&      2.54&      2.69&      2.83&      2.43      2.42&      1.98&      1.52	 & 	0.75 & 1.25 								&0.67(0.82)\tablenotemark{a}	 	&0.07(0.07)&0.71(0.72)\\		
MCG+12--02--001*	&3.07&      3.74&      3.31&      2.69&      2.36&      2.00&      1.38&      1.69&      --&      1.18	&1.05&2.59						&1.60(1.60)		&0.10(0.11)&2.08(2.06) \\	
IC 1623			&4.25&      5.40&      3.67&      3.41&      2.65&      2.29&      1.78&     0.31&     0.86&      1.91 &2.40&3.83								&0.87(1.29)\tablenotemark{a} 	&0.08(0.08)&2.26(2.27) \\ 		
NGC 1068*		&25.24&      24.27&      24.27&      24.63&      25.31&      18.00&      16.52&      16.27&      14.24&      8.37	&16.74&34.94							&5.70(4.88)		&0.27(0.27)&3.09(2.96) \\
NGC 1365*		&41.73&      41.92&      36.75&      26.15&      24.06&      16.17&      10.19&      8.48&      4.65&      1.75&19.62&31.84						&1.37(1.40)		&0.11(0.12)&4.33(4.29)\\	
NGC 1614		&2.61&      3.56&      3.21&      3.43&      3.42&      2.52&      2.38&      1.29&     0.98&     0.85&1.10&2.67			&2.24(2.29)		&0.19(0.19)&2.46(2.36) \\		
IRAS F05189--2524	&--&      0.74&      0.96&      0.87&      1.30&      0.82&     1.54&      1.20&      1.10&      1.01&0.25&0.61						&0.10(0.15)				&0.01(0.01)&0.17(0.15) \\			
NGC 2146*		&17.03&      19.39&      20.68&      17.77&      14.76&      12.77&      10.82&      7.42&      2.52&      5.63	&5.68&15.32						&2.78(2.74)				&0.43(0.43)&7.65(7.60) \\	
NGC 2623		&1.92&      2.18&      2.55&      2.84&      3.34&      3.06&      3.10&      2.89&      1.91&      2.22&0.86&2.44			&0.40(0.71)\tablenotemark{a} 	&0.08(0.08)&0.60(0.61) \\		
NGC 3256*		&17.66&      19.22&      22.71&      18.93&      19.68&      14.30&      11.82&      9.25&      4.80&      4.54&7.87&15.97								&5.11(5.16)		&0.36(0.35)&5.53(5.47)\\		
Arp 299*		&11.26&      9.82&      13.13&      12.56&      12.81&      15.21&      17.27&      15.45&      14.16&      17.67	&2.78&16.05							&5.90(5.93)		&0.58(0.57)&9.13(9.02)\\	
ESO 320--G030		&4.39&      4.56&      5.17&      4.54&      3.69&      4.07&      3.97&      2.48&      2.15&      1.19&	1.72&3.42							&0.77(2.87)\tablenotemark{a}	&0.06(0.06)&1.63(1.66) \\		
NGC 4418		&1.17&      5.57&      6.42&      5.17&      6.08&      6.27&      5.62&      7.05&      5.99&      6.23 	&1.82&1.08	 						&-0.07(-0.13)\tablenotemark{b}	&--&0.14(0.14) \\
Mrk 231			&--&      1.67&      2.06&      2.46&      2.62&      2.17&      2.50&      2.41&      2.02&      1.79 &0.58&1.34							&0.14(0.17)		&0.03(0.03)&0.39(0.38) \\		
IRAS 13120--5453	&--&      5.12&      5.08&      6.70&      6.55&      5.27&      5.11&      4.05&      3.07&      2.33 	&2.35&5.39						&1.09(1.57)\tablenotemark{a}	&0.08(0.10)&1.33(1.34)\\
Arp 193        		&3.20&      3.02&      3.58&      3.41&      3.14&      2.59&      2.37&      1.99&      1.00&      0.90	&1.35&3.70					&1.21(2.23)\tablenotemark{b}		&0.12(0.13)&1.52(1.54) \\
NGC 5135*		&3.53&      3.77&      3.97&      3.04&      2.49&      1.95&      1.50&      1.03&      1.02&     0.42&3.36&6.18						&1.18(1.14)		&0.09(0.09)&1.60(1.54)\\		
ESO 173-G015*		&12.12&      17.14&      19.10&      17.52&      16.88&      12.75&      11.81&      11.14&      7.57&      5.77&4.90&14.48								&1.48(1.73)\tablenotemark{a} 	&0.22(0.21)&2.53(2.51) \\		
Mrk 273			&--&      1.89&      2.55&      2.82&      2.96&      2.68&      3.34&      2.30&      1.58&      1.91 &0.51&1.76				&0.39(0.69)\tablenotemark{a} 	&0.07(0.07)&0.74(0.73) \\
Zw 049.057		&--&      1.84&      2.57&      2.99&      2.57&      2.39&      2.06&      1.62&     0.71&     0.78	&0.96&1.41							&0.06(0.07)\tablenotemark{b} 	&0.02(0.02)&0.35(0.35)\\
Arp 220 		&6.77&      9.35&      12.01&      13.39&      14.53&      12.99&      10.24&      10.18&      5.28&      5.97&2.95&7.42						&-5.80(-5.97)\tablenotemark{c}	&-0.03(0.15)&1.16(1.07)\\
NGC 6240		&7.78&      10.88&      13.90&      16.63&      18.29&      16.87&      16.22&      14.07&      12.08&      10.52 	&2.04&9.54						&6.25(5.99)		&0.50(0.46)&3.74(3.57) \\		
IRAS F17207--0014	&--&      2.81&      4.03&      4.47&      5.29&      3.24&      3.76&      4.31&      2.58&      2.02&1.12&2.66				&0.19(1.57)\tablenotemark{a} 	&0.06(0.08)&0.87(0.89) \\
IRAS F18293--3413 	&5.88&      7.19&      7.55&      6.24&      4.34&      3.46&      2.81&      1.54&     0.98&      1.32	&4.25&7.43				&2.56(4.95)\tablenotemark{a} 	&0.23(0.24)&4.82(4.89) \\	
IC 4687			&1.67&      1.89&      2.02&      1.56&      1.16&     0.61&     0.66&     0.46&     0.34&     0.57&0.91&1.57						&1.40(1.44)		& -- &2.69(2.68) \\		
NGC 7469 		&3.24&      4.57&      4.23&      3.74&      3.10&      2.61&      1.91&      1.59&      1.27&      1.19&2.14&4.63					&1.72(1.77)		&0.16(0.15)&2.04(2.02) \\		
NGC 7552		&12.38&      12.62&      12.98&      11.55&      10.73&      6.86&      4.25&      4.26&      1.92&      1.48&5.46&10. 75						&3.90(3.76)		&0.23(0.23)&4.20(4.09) \\		
NGC 7771*		&4.49&      4.83&      4.59&      3.49&      2.26&      2.17&     0.99&      1.59&     0.15&     0.85 &5.86&7.29				&0.60(0.95)\tablenotemark{a} 	&--&1.64(1.68) \\
Mrk 331*		&2.45&      2.49&      2.69&      2.30&      2.10&      1.91&      1.14&      1.14&     0.72&     0.89 &1.34&2.57						&1.00(1.41)\tablenotemark{a}
&0.09(0.08)&1.82(1.84) \\				
\end{tabular}}

\caption{\footnotesize $^a$ Profile shows a partial absorption feature. $^b$ Profile shows inverse P-cygni profile. $^c$ Profile is in full absorption. $^d$ Upper limit. The integrated fluxes observed with \emph{Herschel}/SPIRE in units of 10$^{-17}$ W m$^{-2}$. Flux errors are 16\% for all SPIRE observations of galaxies that are not extended.  For the extended galaxies (denoted with an asterisk in the table) Arp 299, ESO 173-G015, MCG+12--02--001, Mrk 331, NGC 1068, NGC 1365, NGC 2146, NGC 3256, NGC 5135, and NGC 7771, the error is 30\%. The lines observed with \emph{Herschel}/PACS are [OI]$_{63}$, [OI]$_{145}$, and [CII] are in units of 10$^{-15}$ W m$^{-2}$. For the PACS observations, the number in parenthesis is the flux of the best fit gaussian profile. The fluxes 
characterized by a '--' indicate that lines were not in the observed spectral range. Negative numbers are lines that appear in absorption (Arp 220) or with complex profiles, such as inverse P-cygni profiles (NGC 4418 and Zw 049.057). \label{tab:flux}}
\end{sidewaystable*}
\clearpage

\begin{table}
\centering
\resizebox{.8\textwidth}{!}{
\begin{tabular}{lccc|ccc|ccc}
\hline
\hline
Object & CO J=1-0 & Beam ('')  & Ref. & CO J=2-1 & Beam ('')&Ref. &CO J=3-2 & Beam ('') & Ref. \\
\hline
NGC 34			&0.74&48''&H98&3.06&23''&H98&--&&\\
MCG+12--02--001		&--& &&--&& &--&& \\	
IC 1623			&2.60&22''&P12&17.11&52''&I04 &37.5&11''&P12 \\
NGC 1068		&10.82&22''&P12&86.37&14''&P12&196.05&11''& P12\\
NGC 1365		&13.16&55''&G04&18.70&23''&H98&53.90&14''&I14\\
NGC 1614		&0.79&21''&G04 &1.54&22''&K13&--&&\\
IRAS F05189--2524	&0.18&22''&P12&0.96&14''&P12&2.83&11''&P12\\
NGC 2146		&9.53&21''&G04&6.91&12.5''&B93&84.46&21''&M99\\
NGC 2623 		&0.61&22''&P12&2.01&14''&P12&6.94&11''&P12 \\
NGC 3256		&3.30&44''&A95&53.00&22''& A95&--&&\\
Arp 299 		         &2.23&22''&P12&--&&&49.66&11''&P12\\
ESO 320--G030 		&0.68&48''&M90& --& && --&& \\
NGC 4418		&0.50&22''&P12&--&&&11.37&11''&P12 \\
Mrk 231			&0.32&22''&P12&2.32&14''&P12&6.27&11''&P12\\
IRAS13120--5453		&--& &&--& &&--&& \\				
Arp 193 		&0.73&22''&P12&6.37&14''&P12&13.55&11''&P12\\
NGC 5135		&1.45&22''&P12&9.35&14''&P12&22.25&11''&P12\\
ESO 173--G015		&--& &&--& &&--&& \\	
Mrk 273			&0.30&22''&P12&2.00&14''&P12&5.35&11''&P12\\
Zw 049.057		&0.45&22''&P12&4.58&14''&P12&8.06&11''&P12 \\
Arp 220 		&1.58&22''&P12&8.49&14''&P12&41.53&11''&P12\\
NGC 6240		&1.21 &22''&P12&11.17 &14''&P12&36.00 &11''&P12 \\
IRAS F17207--0014	&0.59&22''&P12&5.06&14''&P12&13.22&11''&P12\\
IRAS F18293--3413	&2.23&55''&G04&--&&& -- && \\
IC 4687/6 		&0.42&48''&A07&2.65&23''&A07&--&& \\
NGC 7469		&1.12 &22''&P12&6.72&14''&P12&18.12&11''&P12\\
NGC 7552		&3.10&48''&C92&21.00&22''&A95&26.00&15''&I14\\
NGC 7771		&1.33&55''&S91&--&& &10.12&23''&N05\\
Mrk 331 		&1.27&55''&G04&--& &&--&& \\
\hline
\end{tabular}}
\caption{\footnotesize Ground based CO integrated fluxes from the literature.  All units are in 10$^{-18}$ W m$^{-2}$.  \label{tab:ground} References are as follows: A95=\citet{1995A&A...300..369A},A07= \citet{2007AA...462..575A}, G04= \citet{2004ApJS..152...63G}, H89= \citet{1989ApJ...342..735H},
I04= \citet{2004ApJ...616L..63I}, I14=Israel, F.P., 2014 A\&A to be submitted, K13= \citet{2013A&A...553A..72K},
M90= \citet{1990AA...236..327M}, M99= \citet{1999AA...341..256M}, N05= \citet{2005ApJ...630..269N},
P12= \citet{2012MNRAS.426.2601P}, and references therein, S91= \citet{2013A&A...553A..72K}.}

\end{table}

\clearpage

\subsection{Classification of CO ladders}
\label{sec:alpha}
\begin{figure*}
\epsscale{2.}
\plotone{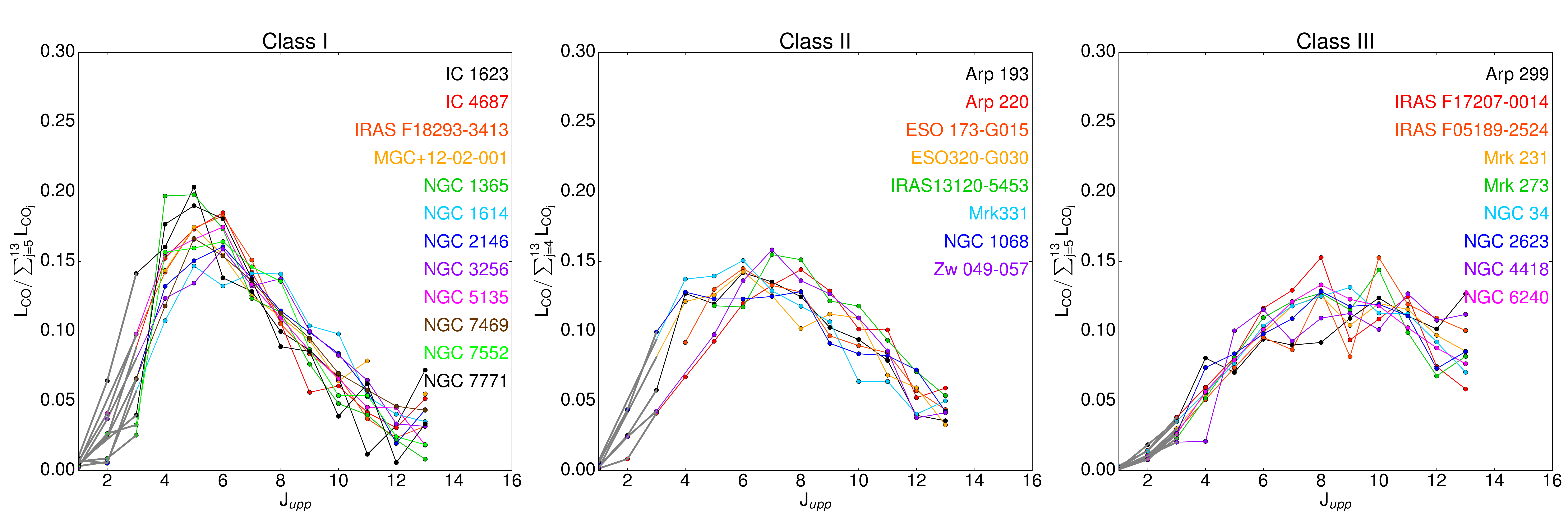}
\caption{CO spectral line energy distributions for the full HerCULES sample divided into three classes.   Class I (left panel) includes galaxies with $\alpha<0.33$, Class II (center panel) 
is where  $0.33<\alpha<0.66$, and 
Class III (right panel) is where $0.66>\alpha$; $\alpha$ is defined in Eq.~\ref{eq:alpha} in Section~\ref{sec:alpha}. Where we do not have data, we have linearly interpolated between the neighboring 
transitions. If we lack all three ground based transitions, we do not plot any low-J fluxes. The ground based transitions have been plotted in gray, to emphasize the \textit{Herschel} line fluxes.}
\label{fig:co}
\end{figure*}

In Figure~\ref{fig:co}, we present the CO ladders of the full HerCULES sample.  We have 
collected the available ground based observations of CO J=1-0, 2-1, and 3-2, whose fluxes and references are listed in Table~\ref{tab:ground}. Where necessary we have converted the ground-based measurements to the cosmology adopted here (Section 1). In 
order to compare these CO ladders directly, we have normalized the ladders by the integrated CO flux summed from J=4-3 through J=13-12, to focus on
the relative behavior of the higher-J transitions since we do not have CO J=1-0 data for all sources.  For galaxies with non detections for any CO transitions, we use
the linearly interpolated value. The 
CO ladders are separated into
three classes based on the parameter $\alpha$, where:
\begin{equation}
\label{eq:alpha}
\alpha=\frac{L_{CO_J=11-10}+L_{CO_J=12-11}+L_{CO_J=13-12}}{L_{CO_J=5-4}+L_{CO_J=6-5}+L_{CO_J=7-6}}.
\end{equation}  
and L$_{CO}=4\pi D_L^2 F_{CO}$, with F$_{CO}$ in [W m$^{-2}$] (Table~\ref{tab:flux}) and luminosity distance, D$_L^2$ in [m] (listed in Table~\ref{tab:obs}).  Here we use three transitions of both the mid- and high-J CO transitions to help prevent noise or a non-detection of one of these lines from dominating $\alpha$.  We 
define the three classes as:
\begin{itemize}
\item Class I: $\alpha<0.33$
\item Class II: $0.33<\alpha<0.66$
\item Class III: $\alpha>0.66$
\end{itemize}
The definition of the classes is quantitatively arbitrary, 
but chosen to reflect similarities in the spectral line energy distributions, which is illustrated in Figure~\ref{fig:co}.  In the case that we do not have any observations of the low-J transitions, we do not plot any low-J fluxes. The parameter $\alpha$ is based on the ratio of three high-J 
CO lines to three mid-J CO lines, which essentially defines the drop-off slope of the CO ladder from J=5-4.  Thus, the steepest drop-offs
are in Class I, while the flattest ladders are in Class III.  Class II consists of objects that peak around J=6-5, 
but do not fall off as steeply as those of Class I. Our three example galaxies were selected to fit into these categories, with NGC 7552 as a Class I, Mrk 331 as a Class II, 
and IRAS F170207-0014 as a Class III object.

We note that the CO ladders for many Class II and III (the excited classes) objects have been published.  In the case of all
of these sources, heating mechanisms besides UV heating are required to explain the high-J CO emission, when also considering additional constraints.
Arp 220 \citep{2011ApJ...743...94R}, Arp 299 \citep{2014A&A...568A..90R}, NGC 253 \citep{2014A&A...564A.126R}, and NGC 6240 \citep{2013ApJ...762L..16M} require mechanical heating to reproduce the high-J CO lines, while 
Mrk 231 \citep{2010A&A...518L..42V} and NGC 1068 \citep{2012ApJ...758..108S} require X-rays to directly heat the gas in order to reproduce the observed
molecular emission.  This trend suggests that when dealing with highly excited CO ladders, such as in Class II and especially Class III objects, 
there is an additional heating mechanism necessary to explain the observed molecular emission. We will explore this issue for the full sample in the next section.

\section{Analysis}
\label{sec:fir}
We will combine our PACS and SPIRE observations of all the major neutral gas cooling lines ([OI], [CI], [CII], and CO), with 
ancillary data obtained as part of The Great Observatories All-Sky LIRG Survey (GOALS) \citep{2009PASP..121..559A}, including the major PDR coolant [SiII] at 34.8$\mu$m. We choose to include 
[SiII] since this is a known PDR coolant that has strength on order of [CII], thus is an important element of the cooling budget.  We will not be dealing with 
any ionized gas coolants such as [NII], [OIII]. We use the IRAS definition of the 
far infrared flux (FIR) as FIR$=1.26\times 10^{-14}$ (2.58 S$_{60\mu m}$+S$_{100\mu m}$)[W m$^{-2}$], where S$_\nu$ is in units of Jansky [Jy] \citep{1985ApJ...298L...7H}.  We then use the 
luminosity distances (D$_L$), from \citet{2009PASP..121..559A}, to define the far infrared luminosity ($L_{\rm FIR}$), making our calculations directly comparable to \citet{2013ApJS..206....1S}.
When we refer to the CO flux, we use the sum of the line fluxes from CO J=4-3 through J=13-12.

\subsection{Warm gas tracers}
\begin{figure*}
\epsscale{2.}
\plotone{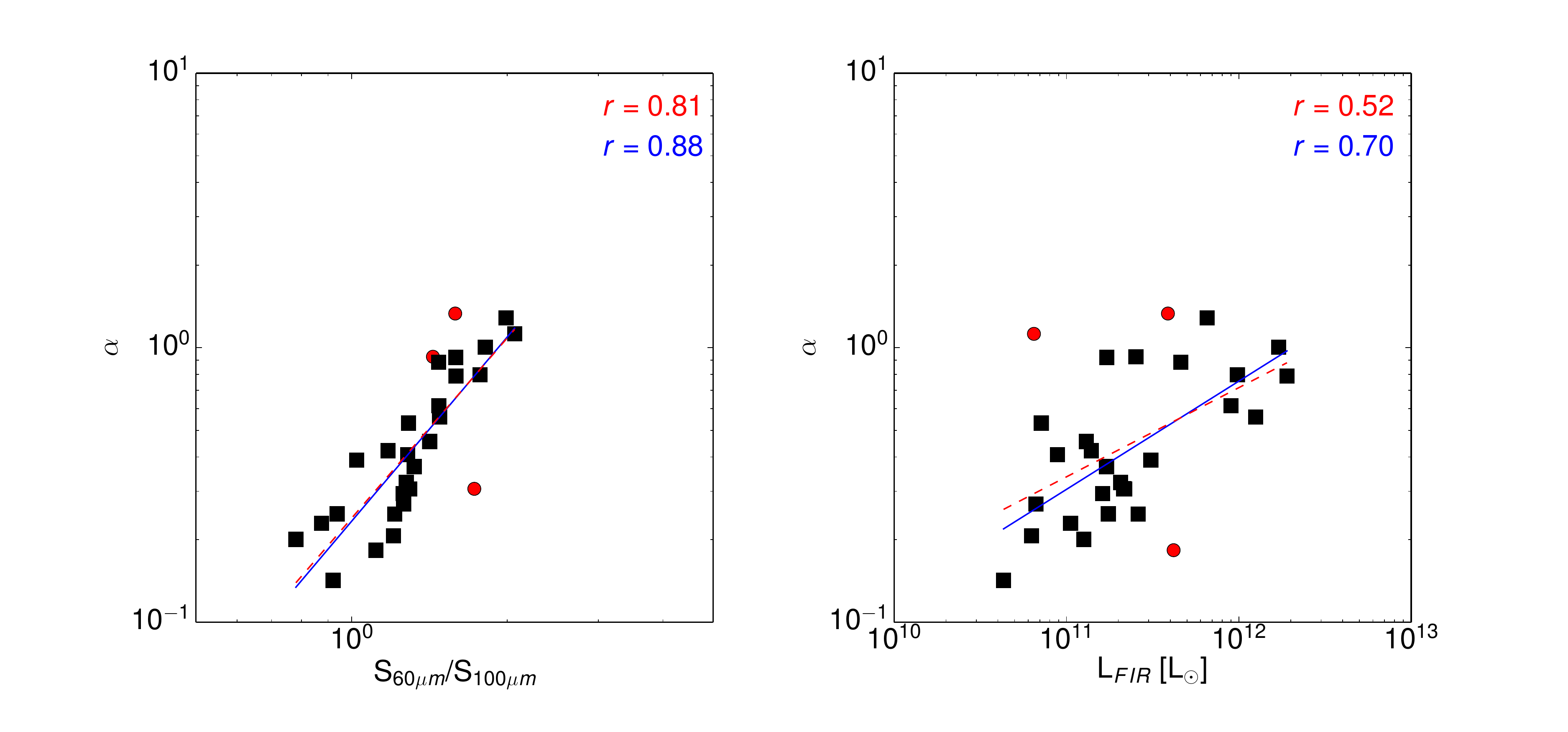}
\caption{Left panel: The gas excitation ($\alpha$, for a definition see Section~\ref{sec:alpha}) plotted against the IRAS infrared color (S$_{60} \mu$m/S$_{100} \mu$m), each square is a galaxy in our sample. The red dashed line and text
represents the least squares fit and correlation coefficient ($r$) for the full sample.  The three red circles represent the the most extreme outliers (largest Euclidean distance value) in the relation
(Arp 299, NGC 1614, and NGC 2623).
The blue solid line and text represent the best fit and correlation coefficient for the sample
excluding the three most extreme outliers, also using the largest Euclidean distance.  Right panel: The same as the left panel but the gas excitation is plotted against the $L_{\rm FIR}$, calculated in Section~\ref{sec:fir}.  The 
three outliers in this case are Arp 299, NGC 4418, and IRAS F18293--3413.}
\label{fig:dusttemp}
\end{figure*}

In the local universe, the ratio of IRAS 60/100 $\mu$m flux densities correlates with infrared luminosity (e.g., \citealt{2003ApJ...588..186C}). It is 
of interest to determine whether the IRAS 60/100 $\mu$m ratio (a proxy for dust temperature) or $L_{\rm FIR}$ (dust luminosity) correlates better with the degree of CO excitation, as parametrized by $\alpha$. 
Since $\alpha$ represents a proxy for the slope of the CO ladder above J=5, it traces the relative brightness of high-J lines in comparison to mid-J lines, allowing a rough estimate of overall CO excitation.  
When $\alpha$ is small, the CO excitation is low and the CO SLED is highly peaked and when $\alpha$ is large, the CO SLED is flattened and the excitation is high, indicating significant emission by warm and dense molecular gas.  We
compare the S$_{60\mu m}$/S$_{100\mu m}$ ratio from \citet{2003AJ....126.1607S} and the $L_{\rm FIR}$ to the molecular gas excitation ($\alpha$).
In the left panel of Figure~\ref{fig:dusttemp} the excitation ($\alpha$) is plotted as a function of S$_{60\mu m}$/S$_{100\mu m}$, where each square point is 
a galaxy in our sample.  The best-fit power law is shown with a red dashed line.  In the
right panel, a similar correlation between $L_{\rm FIR}$ and $\alpha$ is shown, with the best-fit power law plotted with a red dashed line.  We see that although
both panels show a positive trend, the correlation found with the S$_{60\mu m}$/S$_{100\mu m}$ ratio is tighter than that seen with the $L_{\rm FIR}$. The molecular gas 
excitation to infrared color relation has a correlation coefficient of an $r$=0.81, while the excitation to 
$L_{\rm FIR}$ relation has a correlation coefficient of an $r$=0.52.  Although the excitation to infrared color relationship is significantly correlated, we find three 
outliers, based on the largest Euclidean distance and shown in red in the left panel of Figure~\ref{fig:dusttemp}), Arp 299, NGC 1614, and NGC 2623.  We also select the three 
farthest outliers, also based on the Euclidean distance, in the 
$\alpha$ to $L_{\rm FIR}$ relation (Figure~\ref{fig:dusttemp}, Arp 299, NGC 4418 , and IRAS F18293--3413), also plotted as red points in the right panel. We use the traditional definition of 
Euclidean distance: $\left \| u-v \right \|^{2}$. To test the strength of the correlations, we refit a 
power law excluding the three outliers in each plot, which are marked in red.  We remove the outliers to test if the correlation between the CO excitation and the IRAS 
colors is still higher than that of the L$_{FIR}$, or if it is just the outliers affecting the coefficient. The new best-fit is plotted as the blue solid line in Figure~\ref{fig:dusttemp}.  
Although the exclusion of these points does not result 
in a significant change in the best fit in either case, it does improve the correlation coefficients, resulting in a correlation coefficient of an $r$=0.88 for the molecular gas 
excitation to infrared color correlation, and an $r$=0.70 for the molecular gas excitation to $L_{\rm FIR}$ relation. Physically, this suggests that the presence of warm, dense 
molecular gas, is correlated with the presence of warm dust. It is, however,
important to note that, once removing the outliers, the correlation between $\alpha$ and $L_{\rm FIR}$ is also significant, with a 1.6\% probability of this relation being spurious before
removing the outliers. 
Further, since 
the IRAS 60/100 $\mu$m ratio is shown to correlate with $L_{\rm FIR}$, these two quantities are likely related by underlying variables, making this correlation difficult to interpret.
We note however that 
\citet{2014ApJ...787L..23L} compared specific CO 
line transitions normalized by FIR and the IRAS 60/100 $\mu$m flux ratio.  They find that 
as the CO gas becomes warmer, the 60/100 $\mu$m ratio also increases, which is in good agreement with our results. 

\subsection{Cooling budget}
\label{sec:cool}
\begin{figure*}
\epsscale{2.}
\plotone{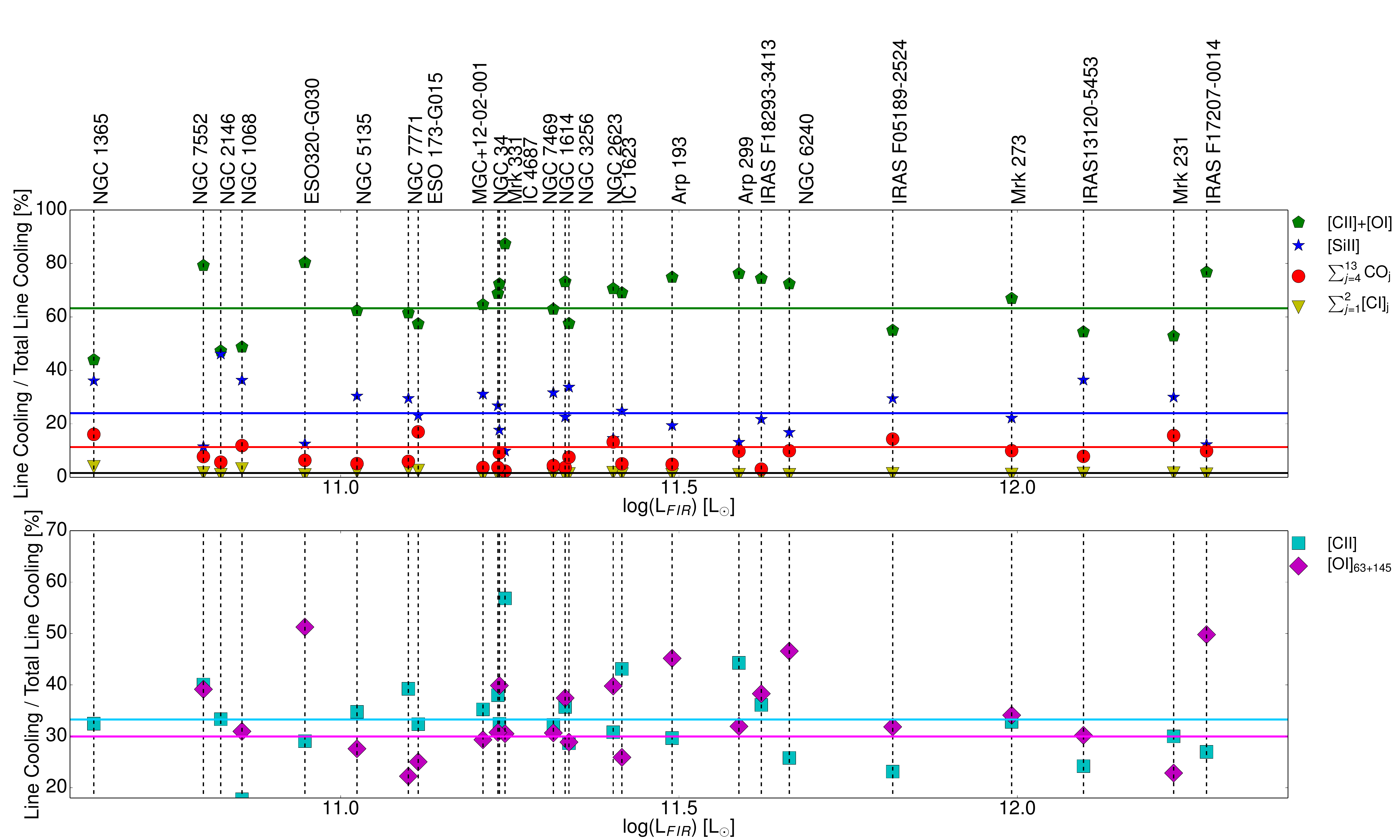}
\caption{\textbf{Top:} Percentage of line cooling of the total gas cooling from each observed species as a function of $L_{\rm FIR}$. Given are absolute, not cumulative percentages.  The yellow triangles are the percentage of cooling 
from [CI]$_{609}$ and [CI]$_{370}$, the red circles are the cooling from CO (4$\leq J_{upp} \leq 13$), the blue stars are the cooling from [SiII], and the green pentagons are from [OI]$_{63+145}$+[CII].  The colored lines represent the mean cooling 
percentages for each coolant. We exclude galaxies that show [OI] in absorption (Arp 220), or 
show a complex line profile (NGC 4418 and Zw 049.057). \textbf{Bottom}: Same as top but separating the cooling contributions of [CII] (cyan squares) and [OI]$_{63+145}$ (magenta diamonds).}
\label{fig:cooling}
\end{figure*}

We can calculate the neutral gas cooling budget in each galaxy by summing the luminosities of the [OI], [CI], [SiII], [CII], and CO lines, since these are the main neutral
gas coolants in the mid- and far-infrared regime.  We take the [SiII] fluxes from \citet{2013ApJ...777..156I}, which were observed with the \emph{Spitzer} IRS instrument in the long wavelength, high resolution mode.  We
note that the slit size for Spitzer IRS-LH is 11.1''$\times$22.3'', thus galaxies that are more extended than 11.1'' will be missing some [SiII] flux.  There are five affected galaxies, Arp 193,
Arp 299, IRAS 13242--5713, NGC 1365, and NGC 2146.  In Figure~\ref{fig:cooling}, we present the percentage of cooling contributed by each emission line as a function of $L_{\rm FIR}$.  The percentage cooling for each species is calculated 
by comparing the luminosity of a species to the total summed luminosity of the [OI], [CI], [SiII], [CII], and CO lines. We exclude three galaxies based on their [OI]$_{63}$ profiles, 
Arp 220 that is fully in absorption, and NGC 4418 and Zw 049.057, which are both heavily absorbed and show inverse P-Cygni profiles \citep{2012A&A...541A...4G}. For each galaxy, the 
percent of cooling contributed by [CI]$_{609}$ and [CI]$_{370}$ is plotted as a yellow circle, CO J=4-3 through J=13-12 in red, [SiII] in blue, and the combined cooling of [OI] 63$\mu$m, [OI] 145$\mu$m, and [CII] is plotted in green.
In the bottom panel of Figure~\ref{fig:cooling}, we separate the cooling contributions of [OI] and [CII].  For the five galaxies that are larger than 11.1'', we expect a higher 
[SiII] contribution than shown in Figure~\ref{fig:cooling}.
The solid lines show the mean percent-cooling for each emitting species.  For example, the neutral atomic carbon is responsible for no more 
than 2\% of the total cooling with an average cooling contribution of 1.5\%, while CO contributes a mean of 10.8\%, and [SiII] contributes 24.2\%. There are two galaxies with exceedingly high 
CO cooling percentages, namely IRAS F05189--2524 and Mrk 231, the two strongest AGN in the sample.  Very high CO cooling percentages have also been noted 
in the massive Galactic star forming region W3 with 32\% total gas cooling \citep{2004A&A...424..887K}, and even higher percentages in DR21 \citep{2007A&A...461..999J,2010A&A...518L.114W}.  In both cases, 
the high CO percentage is attributed to self absorption of the [OI]$_{63}$ line, yet we calculate our fluxes both with the Gaussian and observed [OI]$_{63}$ line fluxes and see little change. Therefore,
for IRAS F05189--2524 and Mrk 231, we believe the high CO cooling percentage is a true effect, and not one dependent on the [OI]$_{63}$ absorption. The most efficient
coolants are [OI] and [CII], which together provide a mean of 63.7\% of the total gas cooling budget. Separating their cooling contributions, [CII] provides a
mean cooling percentage of 33.6\% and [OI] cools a mean of 30.1\% of the gas. The mean cooling 
percentages and their standard deviations are shown in Table~\ref{tab:pcooling}. Inspection of
Figure~\ref{fig:cooling} shows that the outliers are randomly distributed, and there is no clear trend of 
outliers as a function of $L_{\rm FIR}$.

\begin{table}[h]
\centering
\caption{Mean and standard deviation of percent cooling contribution.}
\setlength{\tabcolsep}{4pt}
\begin{tabular}{|l|c|c|}
\hline
\hline
Line & Mean & Std. Dev.\\
\hline
$\rm [CII] +[OI]_{63+145}$ &63.7 &14.3  \\
$\rm[CII]$                 &33.6 &9.1  \\
$\rm [OI]_{63+145}$        &30.1 &11.8 \\
$\rm [SiII]$		   &24.2 &9.9     \\
$\sum_{j=4}^{13}\rm CO_j$  &10.8 &10.0  \\
$\sum_{j=1}^{2}\rm [CI]_j$ &1.5  &0.9  \\
\hline
\end{tabular}
\label{tab:pcooling}
\end{table}

\begin{figure*}
\epsscale{2.}
\plotone{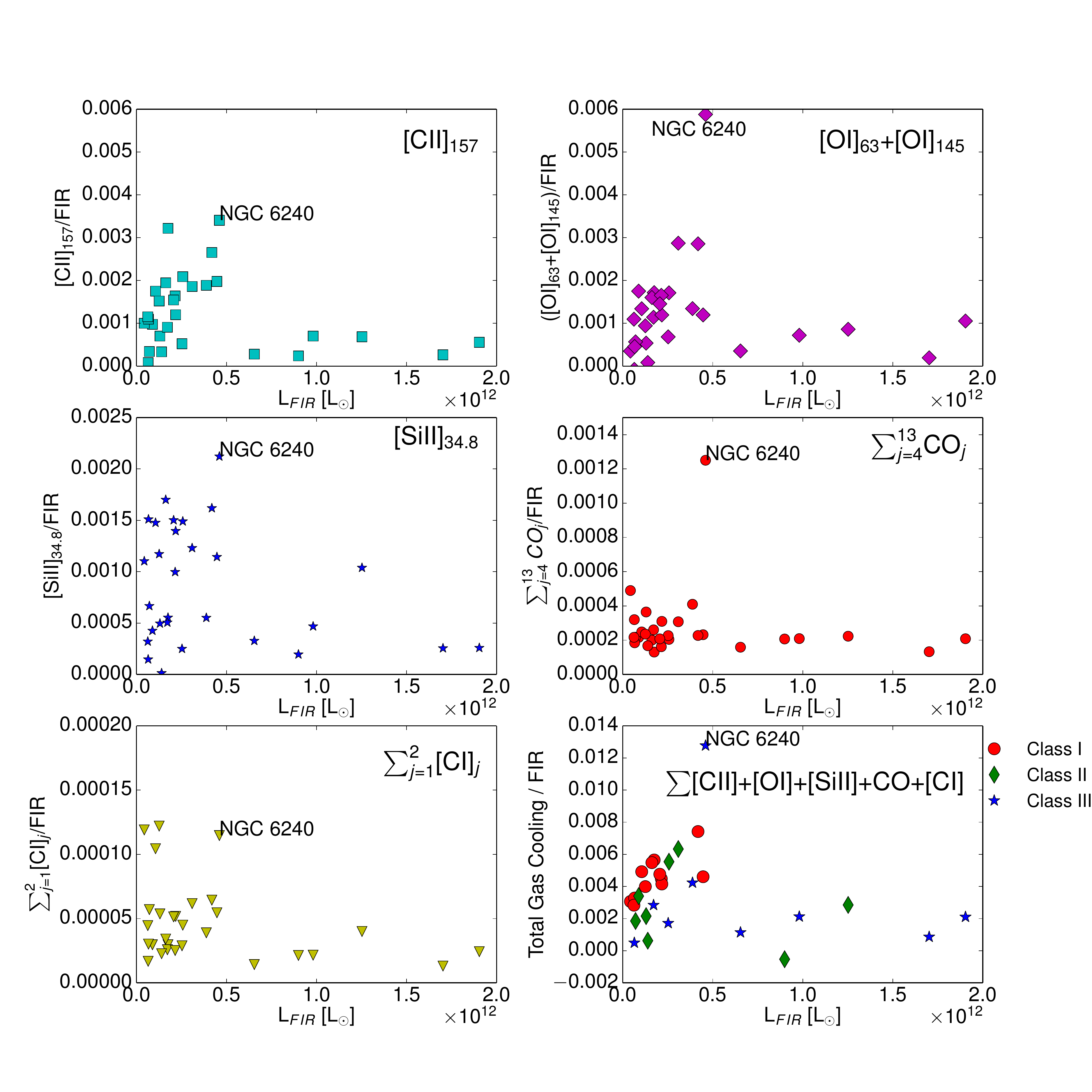}
\caption{The ratio of line fluxes to far infrared flux (FIR) plotted as a function of far infrared luminosity ($L_{\rm FIR}$) (both defined in Section~\ref{sec:fir}) for [CII] (top left), 
[OI] (top right), [SiII] (middle left), CO (middle right), [CI] (bottom left), and the total cooling (bottom right).  The [OI] flux is derived from the sum of [OI]$_{63}$ and [OI]$_{145}$, 
the CO flux is the sum of line transitions from J=4-3 to J=13-12, and the [CI] flux is the sum of [CI]$_{609}$ and [CI]$_{370}$. The total cooling refers to the total neutral gas cooling 
and includes all aforementioned fluxes. We 
exclude Arp 220, NGC 4418, and Zw 049.057 from the [OI] deficit and total neutral gas cooling deficit plots, since we only observe these lines in full or partial absorption and can thus not get an accurate flux estimate. }
\label{fig:def}
\end{figure*}


It is interesting to note how constant each cooling range is as a function of $L_{\rm FIR}$.  It is a well known phenomenon that as far infrared 
luminosity increases, an apparent [CII]/FIR ratio decreases, which is what we call the [CII] deficit.  This is observed in various environments in the local Universe \citep{2001ApJ...561..766M,2003ApJ...594..758L,2013ApJ...774...68D}.  In addition,
\citet{2011ApJ...728L...7G} find a fine-structure line deficit in [NII] and [OI] as well, such that both the [NII]/FIR and [OI]/FIR ratios decrease as a function of far infrared luminosity. One corollary to this fine-structure line deficit is 
that other (molecular) coolants could become more efficient at higher $L_{\rm FIR}$.  However, as shown in Figure~\ref{fig:cooling}, the relative efficiencies of each coolant remain mostly constant. 
In order to better understand this phenomenon, we plot the [CII] deficit ([CII] to FIR ratio), as a function of $L_{\rm FIR}$ in the top left corner of Figure~\ref{fig:def}.  
For comparison, we plot the 
same ratios but for the [OI]$_{63+145}$, [SiII], CO, and [CI] lines, where the CO lines encompass all transitions 4$\leq J_{upp} \leq 13$ and [CI] is the sum of both [CI]$_{609}$ 
and [CI]$_{370}$.  We see a trend that is consistent with a deficit in [CII], agreeing with the observations from \citet{2013ApJ...774...68D}.  Since our sample spans a smaller luminosity 
range than
\citet{2013ApJ...774...68D}, our deficit only spans a factor of 2 in the [CII]/FIR ratio, while in the GOALS sample, it spans an order of magnitude.  Thus, we plot the deficit in linear space and see clearly 
that the only points with a high [CII]/FIR ratio are those of lower $L_{\rm FIR}$.  We see a similar trend in [SiII] and [CI], where the latter has also been observed from ground based observatories \citep{1998ApJ...509L..17G,2000ApJ...537..644G}.
The [OI] shows a tentative line deficit, which becomes more obvious when we include galaxies of higher IR luminosities \citep{2011ApJ...728L...7G}.
Although we find evidence consistent with line deficits in [CII], [OI], [SiII], and [CI], there is no evidence for a deficit in CO, which shows a flat distribution over $L_{\rm FIR}$, further strengthening the results 
from \citet{2014ApJ...787L..23L}.  In all panels, NGC 6240 is an extreme outlier, as was also noted in \citet{2014ApJ...787L..23L}.

The difference between the molecular and fine structure emission can 
be understood in terms of heating mechanisms.  The fine structure lines are originating from regions that are heavily affected by UV photons, at the edges of PDRs.  As shown in the 
PDR models from \citet{1999ApJ...527..795K}, as the
radiation field and density increase, the fine structure line emission is expected to weaken compared to the far infrared flux. However, this 
does not apply for the molecular gas (CO), where we see no line deficit.  This result may not be surprising since CO traces the molecular gas deeper in molecular clouds, where the UV field is significantly attenuated. We can 
test the gas versus dust cooling efficiency by plotting the ratio of the total gas cooling to the far infrared flux as a function of $L_{\rm FIR}$, which is shown in the bottom right panel of Figure~\ref{fig:def}.  Here, 
it is clear that the trend is decreasing in a very similar manner to that of the fine structure lines, which is expected since [CII],[OI], and [SiII] dominate the cooling.  Therefore, we can say that the fine structure line deficit is actually a 
gas cooling deficit in comparison to the FIR flux that is a result of UV heating becoming a more efficient coolant of dust in exceedingly extreme environments. It is 
critical to note that the fine structure line fluxes are not decreasing in absolute flux, but their relative contribution in comparison to the warm dust (measured with the FIR flux) is decreasing, due 
to the warm dust becoming increasingly efficient at a faster rate than the gas coolants. This can also be understood as dust being heated more efficiently, such that the fine structure lines do not actually show a deficit,
which agrees with the results of \citet{2013ApJ...774...68D}, where they find that grains are 'stealing' photons 
from the gas.  

However, if the neutral gas cooling efficiency is decreasing, and we observe a tentative line deficit in all species except for CO, then we would expect the percentage of CO cooling to increase 
slightly as a function of $L_{\rm FIR}$, to compensate for the decreasing fine structure line cooling efficiency.  We do see a slight increase in the percentage of CO cooling, but the trend is 
tentative and within the general scatter of the other CO cooling percentages. In addition, the galaxies with high $L_{\rm FIR}$ have flatter CO ladders, meaning there is 
non-negligible flux in high-J (J$>13$) transitions, that we do not account for in our cooling budget. This missed high-J flux would increase the CO cooling percentage, and could make the 
CO cooling budget increase for higher luminosity sources. Since we only have four galaxies in our sample within the range that we would expect to 
see an elevated CO cooling percentage, we would need to increase our sample size in the high luminosity regime to determine if this trend is real.



These results have several implications. Firstly, it is unlikely that the [CII] line deficit could be caused by high dust opacities at 158 $\mu$m, since the line deficit 
is also observed in lines of much longer wavelengths, notably the [CI] line. Similarly, the line deficits are not likely caused by the line being optically thick, since the 
deficit is also observed in the optically thin [CI] lines, which agrees with the results of \citet{2013ApJ...774...68D}. A likely explanation is star formation dominated by ultracompact HII regions, which 
would suppress all fine-structure lines due to an increase of dust competition for UV photons.  This is further supported by the fact that the total gas cooling 
([CII]+[OI]+CO+[CI]) decreases as a function of $L_{\rm FIR}$, meaning that dust becomes an even more efficient coolant as the infrared luminosity increases.

The lack of a strong CO-line deficit shows that the bulk of the molecular gas heating is not affected by the mechanism suppressing the fine structure lines. 
This is interesting, since in a UV-photon heated environment, suppressing the UV field implies a reduced heating rate, and therefore also a lack of warm molecular 
gas might be expected. We also note that the integrated CO luminosity, that represents 10s of percent of the total gas cooling, is much more than what is predicted by any 
pure PDR model, which give a CO cooling fraction of at most a few percent (3-5\%) of the total gas cooling (e.g., \citealt{2011A&A...525A.119M}). This result suggests that 
the CO luminosity may be powered by a different heating mechanism, which does not lead to dissociation or ionization.

\subsection{Heating Mechanisms}
The CO molecule can be heated indirectly through (a) the photoelectric effect by ultraviolet (UV) photons, (b) by fast electrons from directly ionized H and H$_2$ by X-rays, cosmic rays 
\citep{2005A&A...436..397M}, or (c) mechanical processes, 
which includes shocks and turbulence. X-rays heat gas
by ionizing H and H$_2$ directly, and these fast electrons then thermalize the molecular gas with an efficiency of 10\%.  UV photons ionize PAHs and dust grains, and the resulting free electrons heat the molecular gas with a net efficiency of 1-3\%.  In addition, the chemistry in an XDR is driven by X-ray photons
instead of FUV photons that are able to penetrate deeper into the cloud without efficiently heating the dust at the same time.  These X-rays are
mostly produced by active galactic nuclei (AGN) or in areas of extreme massive star formation. Cosmic rays can also heat the gas by penetrating into
cloud centers, similarly to X-rays, and are typically produced by supernovae.  Mechanical heating is another efficient source of gas heating. This is 
commonly attributed to turbulence in the ISM, which may be driven by supernovae, strong stellar winds, jets, galaxy mergers, cloud-cloud shocks, shear in the gaseous disk, or outflows.

\begin{figure}
\epsscale{1.}
\plotone{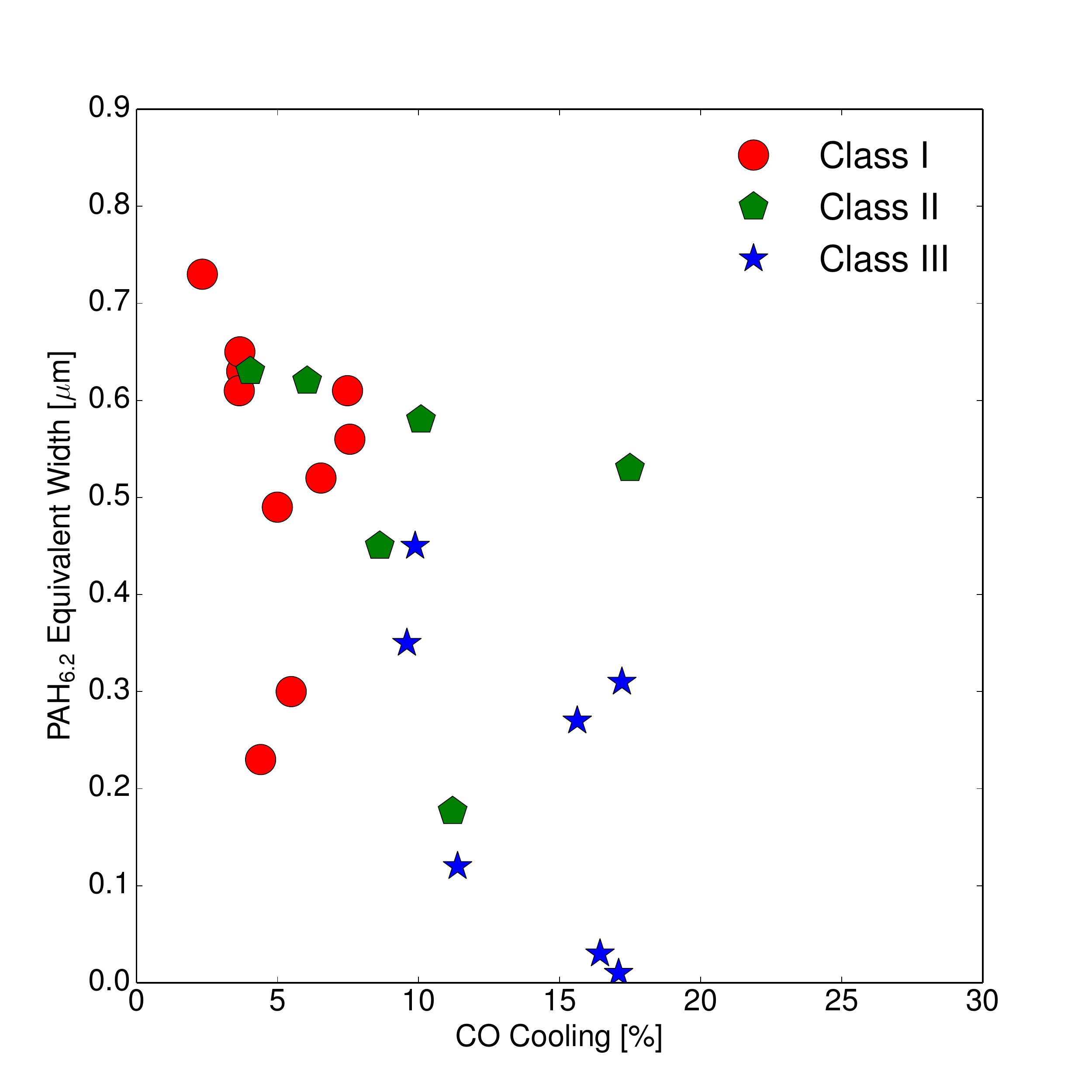}
\caption{The PAH 6.2 $\mu$m equivalent width from the Spitzer IRS \citep{2013ApJS..206....1S} as a function of percentage of CO cooling (for $J>4$), as calculated in Section~\ref{sec:cool}.  
Each galaxy is color-coded by class, red is Class I, green is Class II, and blue is Class III. In addition, we exclude Arp 220, NGC 4418, and Zw 049.057 since their [OI] profiles are fully
in absorption or show a complex line profile, making our CO cooling percentage inaccurate.  Finally, we exclude NGC 1365 since it is extended and we do not capture 
all of the CO emission in one SPIRE pointing, affecting the accuracy of our CO cooling percentage.}
\label{fig:pah}
\end{figure}

To investigate the main mechanism heating the molecular gas, we use another diagnostic molecule, namely polycyclic aromatic hydrocarbons (PAHs). 
PAHs are carbonaceous, nanometer sized macromolecules that contain 50-100 carbon atoms with an
abundance of 10$^{-7}$ per hydrogen atom \citep{2008ARA&A..46..289T}.  The absorption of one far-UV photon is enough to heat the
PAH molecule to a high temperature and will cause this molecule to emit in the characteristic bands at 3.3, 
6.2, 7.7, 8.6, and 11.2 $\mu$m (\citealt{2008ARA&A..46..289T}, and references therein).  Because PAHs are only 
fluorescently excited, and are easily destroyed by more energetic radiation, they are ideal tracers of 
UV heating.  Thus, we can use the equivalent width of the 6.2 $\mu$m feature from the \emph{Spitzer} Space Telescope \citep{2013ApJS..206....1S}, as a 
proxy for the UV energy density. In Figure~\ref{fig:pah}, we compare the PAH equivalent width to the percentage of the total gas cooling done
by CO, as calculated in Section~\ref{sec:cool}. Class I objects have the steepest decreasing CO SLED,  high PAH equivalent
widths (EW), and low percentages of CO cooling. On the other hand, Class II and Class III objects with high percentages of CO cooling have low PAH equivalent 
widths. We note that the two objects with very low PAH equivalent widths are the most AGN-dominated objects in our sample (Mrk 231 and IRAS F05189--2524), where PAH destruction by X-rays 
and nuclear hot dust combine to strongly lower the PAH equivalent width. The high CO cooling fractions of these objects can likewise be attributed to energy input by the AGN (through X-ray 
heating or mechanical energy input from an AGN-driven outflow). There are two galaxies that are Class I objects with lower PAH equivalent widths, and those are IC 1623 and NGC 7469.  
NGC 7469 is a starbursting galaxy with a Seyfert 1 nucleus and IC 1623 is a late-stage merger with a starburst nucleus. NGC 7469 has a central starburst ring along with its AGN nucleus
(e.g. \citealt{2007ApJ...671.1388D}).  This suggests that the Class I type CO emission is coming from 
the starburst ring that is encircling, but not directly affected by the AGN. These trends reinforce the idea that objects with a
high CO cooling fraction, the CO is efficiently excited by something besides UV photons.

\begin{figure}
\epsscale{1.}
\plotone{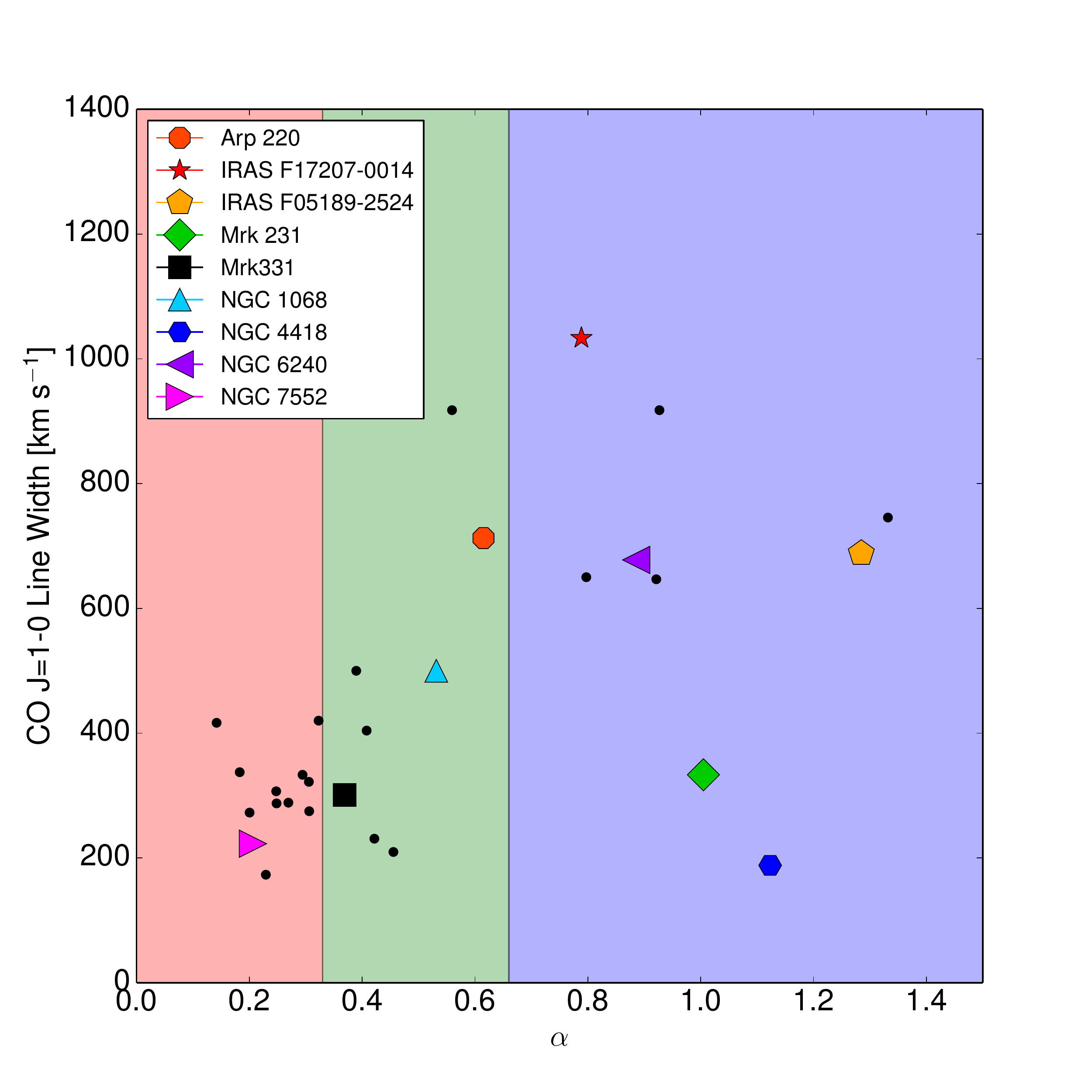}
\caption{CO J=1-0 linewidth (FWHM) in km s$^{-1}$ plotted against $\alpha$ (Eq.~\ref{eq:alpha}). The different classes are highlighted with different background colors, red is Class 
I, green is Class II, blue is Class III.  Most sample galaxies are plotted with black dots, but some of those specifically mentioned 
in the discussion are highlighted.  This also includes the three targets addressed in Figures 1-3.}
\label{fig:lw}
\end{figure}

\begin{figure}[h]
\epsscale{1.}
\plotone{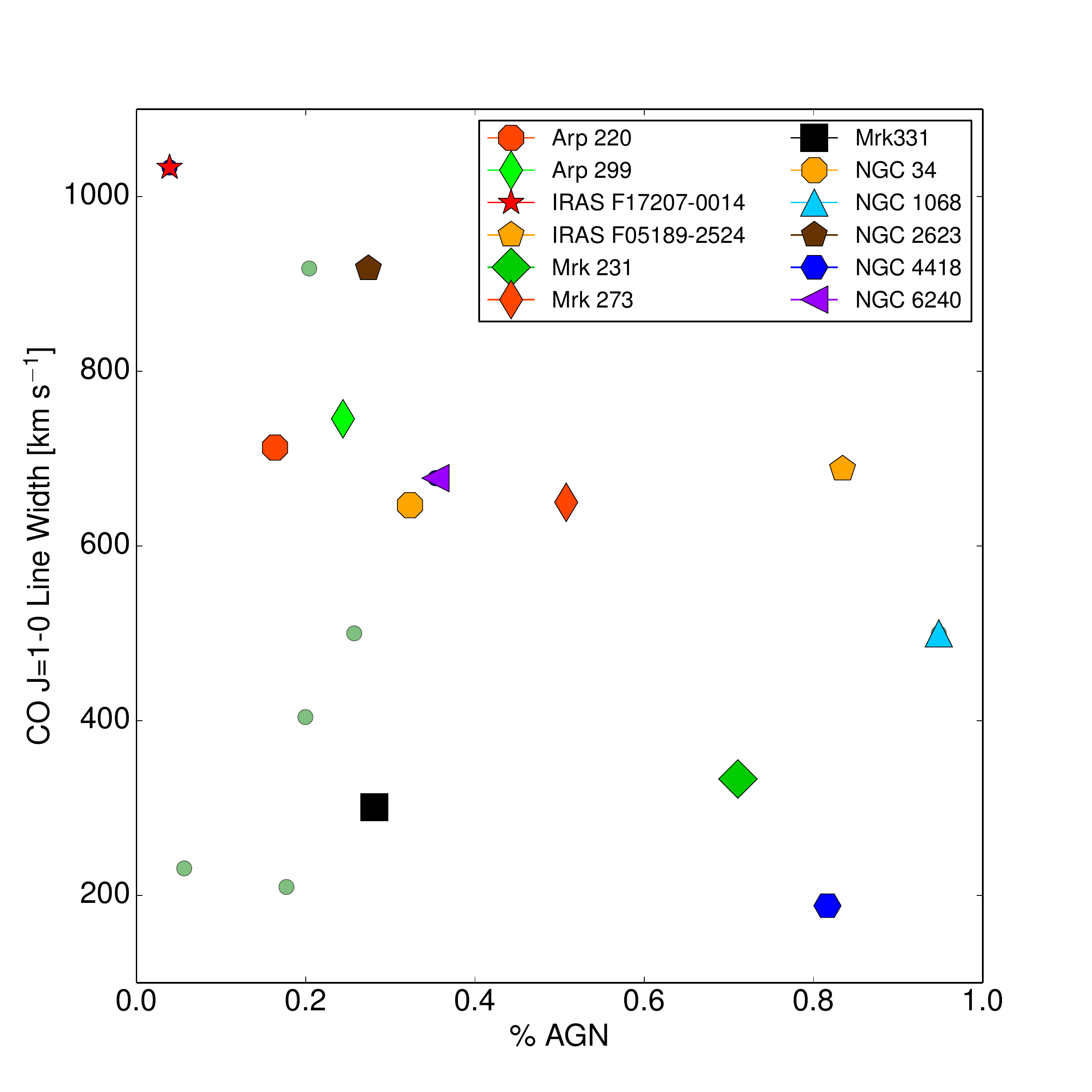}
\caption{The AGN contribution of the bolometric luminosity compared to the CO J=1-0 inclination corrected linewidth for Class III galaxies along with the Class II template galaxies presented in 
Figure~\ref{fig:lw}.  We have also included the rest of the Class II galaxies as transparent green points for reference.}
\label{fig:lw1}
\end{figure}

We now compare our $\alpha$ parameter with the CO J=1-0 linewidth for the sources in our sample (Figure~\ref{fig:lw}). We correct the linewidth for inclination such that all galaxies are effectively turned edge on
using the K-band axis ratio from 2MASS, with the exception of Arp 220 and NGC 6240.  For Arp 220 we used the inclination derived by \citet{1997ApJ...484..702S} using arcsecond imaging of 
CO J=1-0 and for NGC 6240 we used the inclination derived by \citet{2010A&A...524A..56E} using subarcsecond near-infrared imaging. The range of Class I objects is highlighted
with a red, Class II with a green, and Class III with a blue background.  

Merging and interaction, molecular outflows, and random motions may all play a role in increasing the 
linewidths, but are not expected to dominate the linewidths of our target sources.  The linewidth of CO J=1-0 is dominated by rotation, which is determined 
by the mass of the central regions of the galaxy. Therefore, galaxies with high excitation ($\alpha$) and high linewidths are more massive. However, the speed of 
rotation is also a proxy for the mechanical energy reservoir available in the galaxy nucleus since rotation leads to processes such as sheering and turbulence. If the fraction of 
mechanical energy that is converted into heating the molecular gas is constant, then high linewidth galaxies in Figure~\ref{fig:lw} would have more mechanical energy available, and
would be responsible for heating more of the molecular gas. However, establishing this result would require a detailed study of the velocity 
fields of the molecular gas in all of our targets, now possible with ALMA. Since such data is not available, thus we suggest that the linewidth provides an estimate of the available reservoir of mechanical energy, 
but we remain agnostic as to the extent that this reservoir is actually tapped.

Inspecting Figure~\ref{fig:lw} with these considerations in mind, the fact that low-excitation galaxies all have the smallest linewidths, while high-excitation galaxies have a range of linewidths has 
an attractive physical interpretation. In high excitation galaxies with high linewidths, mechanical energy input is a viable source of excitation, while in high excitation galaxies with low 
linewidths, radiative energy input may be more important. Indeed, in Fig. 10, the lowest linewidths among high excitation galaxies are found in Mrk 231 and NGC 4418 (an exposed and an obscured 
AGN, respectively).
Class I and II objects with low linewidths are likely dominated by UV heating, since their 
CO ladders turn around somewhere before J=7-6.  For comparison, we 
have highlighted the position of our three example galaxies, NGC 7552 (Class I), Mrk 331 (Class II), and IRAS F17207-0014 (Class III), 
as well as some of the more famous galaxies in our sample.  The major merger galaxies NGC 6240 and Arp 220 have a relatively high $\alpha$ values and linewidths around 650 km/s.  On the other hand, the strongest AGN in our sample, Mrk 231, lies in the low linewidth region of Class III, suggesting
that its gas is radiatively heated, likely by X-rays from the AGN.  This confirms the results of \citet{2010A&A...518L..42V}, where they find the high-J CO excitation consistent with 
an XDR.  The other two galaxies with AGN contribution are NGC 1068 and 
IRAS F05189--2524, both of which are Class II objects that lie in the low linewidth region, suggesting that they are also radiatively excited.  However, 
since both of these objects lie in Class II (albeit on the border between Class II and III), it is unclear whether their excitation is 
from UV, higher energy photons, or cosmic rays.  Both NGC 1068 and IRAS F05189--2524 have also been studied in detail by \citet{2012ApJ...758..108S} and \citet{2014arXiv1404.6470P} respectively.
For NGC 1068, \citet{2012ApJ...758..108S} and \citet{2014arXiv1405.7706G} find that indeed, the excitation is due to either XDRs (in the circumnuclear disk) or PDRs (in the star-forming ring), which agree 
with our results.  In IRAS F05189--2524, \citet{2014arXiv1404.6470P} find that there is a large contribution from mechanical heating in this source, as traced by a shallow H$_2$ temperature
distribution, yet they cannot rule out the AGN as an important heating source for the molecular gas.

In order to examine this method vis-a-vis the role of AGNs and to determine additional heating sources, and to check if there are any underlying biases in the J=1-0 linewidth, we can compare the 
inclination corrected linewidths with the percentage AGN contribution for each galaxy.  In the case of very disturbed major mergers, such as Arp 220, Arp 299, IRAS 17208--0014, and NGC 6240, 
the inclination is difficult to determine, and thus the corrected linewidth can be over- or underestimated.  The uncorrected linewidths are shown in Table~1 for comparison. 
The AGN contribution can 
be estimated using both the 15/30 $\mu$m flux density ratio ($f15/f30$)and the [NeV]/[NeII] ratio along with the prescriptions from \citet{2009ApJS..182..628V}.  We take the $f15/f30$ ratio from \citet{2013ApJS..206....1S}
and the [NeV]/[NeII] ratio from \citet{2013ApJ...777..156I}.  We average the results of the AGN contribution from both methods and find average
AGN contributions ranging from to 0\%-95\% of the bolometric luminosity. For the high excitation sources (Class III) we use the average AGN contribution in combination with
the inclination corrected linewidth to separate mechanical heating from AGN heating.  In Figure~\ref{fig:lw1}, we plot the AGN contribution on the x-axis and the inclination corrected
CO J=1-0 linewidth along the y-axis.  This scatter plot shows that for our high-excitation galaxies, very high linewidths are not associated with high AGN contributions and conversely 
 the galaxies with high AGN contributions do not display high linewidths. We also plot the Class II galaxies as green points (the template galaxies are marked) for comparison.  Most Class II 
 galaxies have low AGN contributions, while they have a large range of linewidths. 

%


For example, we can compare Arp 220 and Mrk 231.  Both galaxies have high $\alpha$ values, but Arp 220 has a high linewidth and a low AGN contribution, while Mrk 231 has a low linewidth and a 
high AGN contribution.  Comparing NGC 6240 and Arp 220 reveals that although both have high linewidths and $\alpha$ values, NGC 6240 also has a high AGN contribution.  This suggests that 
both the AGN and mechanical processes are contributing to heating the gas in NGC 6240. 

It is difficult to conclude anything definitive about objects that show average or typical values of 
$\alpha$ or linewidths. It is also important to note that although there is a non-negligible contribution of heating from mechanical processes or the AGN in Class III galaxies, the gas is still
mostly heated through UV heating processes.  We caution the use of any two of these diagnostics alone, since for example, a Class III object with a low linewidth 
may still be mechanically heated by small-scale turbulence that would not produce an observable global line broadening effect. In this case, the AGN contribution would be low, but the 
$\alpha$ would be high, discounting AGN heating.  Using all three parameters simultaneously allows for a qualitative estimate of which additional processes are 
exciting the warmest molecular gas. In order to fully understand the excitation mechanisms and physical parameters of the molecular gas, an additional detailed modeling of the $^{12}$CO, $^{13}$CO, and 
dense gas tracers (HCN, HNC, HCO$^+$, etc.) is required (e.g. \citealt{2014A&A...564A.126R}).

\section{Conclusion}
We report the initial results of the Herschel Open Time Key Project Herschel Comprehensive (U)LIRG Survey (HerCULES).  Both \emph{Herschel}/SPIRE spectra and \emph{Herschel}/PACS [OI]63, [OI]145 and [CII] 
line profiles of a sample of 29 galaxies spanning an order of magnitude of infrared luminosity were analyzed. Our main results are summarized below:
\label{sec:conc}
\begin{itemize}
 \item We separate our sample of luminous galaxies into three qualitative classes based on the shape of their CO ladder, quantized by the parameter $\alpha$, which we define as the ratio of the high-J to mid-J
 CO transitions.  
 
 $\alpha=(L_{CO_J=11-10}+L_{CO_J=12-11}+L_{CO_J=13-12})/(L_{CO_J=5-4}+L_{CO_J=6-5}+L_{CO_J=7-6}$)
 
 Class I ($\alpha<0.33$) is characterized with a CO SLED peak around J=5-4 and a steep decline towards higher J transitions.  
 Class II ($0.33<\alpha<0.66$) has a CO SLED peak around J=7-6 and a shallower decline towards higher J transitions.  Class III 
 ($\alpha>0.66$) shows very flat CO ladders.  We present the spectra of three exemplary galaxies for each of these three categories.

 \item We find that molecular gas excitation (approximated by $\alpha$) is well correlated with the infrared color (as traced by $S_{60}$/$S_{100}$), and not as well
 correlated with the $L_{\rm IR}$. 

 \item The cooling budgets of the galaxies are presented.  We find that the percentage of cooling from each species ([CII], [SiII], [OI], and [CI]) appears to be constant over 
 the full range of $L_{\rm FIR}$.  There is indication of a slight increase in the percentage of CO cooling at higher $L_{\rm FIR}$.  
 
 \item We find [CII]/FIR, [SiII]/FIR, and [CI]/FIR ratios consistent with deficits (i.e., the [CII]/FIR ratio decreases with increasing L$_{FIR}$), and also a weak [OI] deficit for the high luminosity galaxies.  On the other 
 hand, we observe no CO deficit, the CO/FIR ratio is very constant for all $L_{\rm FIR}$, with the exception of NGC 6240.  Thus, the fine structure line deficits reflect a decrease in the total gas heating efficiency 
 with increasing L$_{FIR}$.  The fact that we observe a deficit in all fine structure lines but not in the molecular gas suggests 
 that the mechanism responsible for heating the [CII], [OI], [SiII], and [CI], is not the same mechanism responsible for heating the CO. The CO may instead be affected by a heating mechanism that is immune to this deficit. We also
 find that the total neutral gas cooling per FIR decreases as a function of L$_{FIR}$. 
 
 \item Using the PAH 6.2 $\mu$m equivalent width as a proxy for the importance of massive star formation and therefore UV excitation, we find that when the cooling efficiency of CO
 is high, the amount of UV heating is low.  This again indicates that CO is more efficiently heated by a mechanism not directly related to UV radiation. 
 
 \item We suggest a qualitative schematic based on $\alpha$, the CO J=1-0 linewidth, and the AGN contribution, that helps indicate which additional mechanism, if any, is heating the gas.  
 Class I galaxies with a low $\alpha$ ($\alpha<0.33$) do not require any heating in addition to UV-heating to explain the observations.  Class III galaxies with high linewidths and low AGN contributions probably
 require mechanical, in addition to UV heating. Class III galaxies with narrow linewidths and large AGN contributions
 are experiencing excitation from harder radiation (X-rays or cosmic rays).  Class III objects with wide linewidths and high AGN contributions are composite galaxies that are 
 being heated by both mechanical processes and the AGN.  For the objects that have median $\alpha$, linewidth, or AGN contribution values, such as many Class II objects, it is not 
 possible to discriminate which heating mechanisms are affecting the gas without additional information. 

\end{itemize}

\acknowledgments
We would like to thank Edward Polehampton for his help reducing the SPIRE observations.  Basic research in infrared astronomy at the Naval Research Laboratory is funded by the Office of Naval Research.  
JF also acknowledges partial support from the NHSC/JPL subcontract 1371112.SPIRE has been developed by a consortium of institutes led by
Cardiff Univ. (UK) and including: Univ. Lethbridge (Canada);
NAOC (China); CEA, LAM (France); IFSI, Univ. Padua (Italy);
IAC (Spain); Stockholm Observatory (Sweden); Imperial College London, RAL, UCL-MSSL, UKATC, Univ. Sussex (UK);
and Caltech, JPL, NHSC, Univ. Colorado (USA). This development has been supported by national funding agencies:
CSA (Canada); NAOC (China); CEA, CNES, CNRS (France);
ASI (Italy); MCINN (Spain); SNSB (Sweden); STFC, UKSA
(UK); and NASA (USA). The Herschel spacecraft was designed, built, tested, and 
launched under a contract to ESA managed by the Herschel/Planck Project team by 
an industrial consortium under the overall responsibility of the prime contractor 
Thales Alenia Space (Cannes), and including Astrium (Friedrichshafen) responsible 
for the payload module and for system testing at spacecraft level, Thales Alenia 
Space (Turin) responsible for the service module, and Astrium (Toulouse) responsible 
for the telescope, with in excess of a hundred subcontractors. HCSS / HSpot / HIPE is 
a joint development (are joint developments) by the Herschel Science Ground 
Segment Consortium, consisting of ESA, the NASA Herschel Science Center, and 
the HIFI, PACS and SPIRE consortia. HAS acknowledges partial support from NASA Grant NNX12AI55G and JPL RSA contract 717437 and 717353. MHDvdW is supported by the Canadian Space Agency (CSA) and the
Natural Sciences and Engineering Research Council of Canada (NSERC).

%

\bibliographystyle{apj}
\bibliography{bib_file_ulirg}

\end{document}